\documentclass[aps,prl,reprint,superscriptaddress]{revtex4-1}
\usepackage{graphicx}
\usepackage{amsmath}

\usepackage{bm} 
\usepackage{bbold} 
\usepackage{graphicx}
\usepackage{graphics}
\usepackage{fancyhdr}
\usepackage{marvosym}
\usepackage{hyperref}
\usepackage{amsmath}
\usepackage{slashed} 
\usepackage{latexsym}
\usepackage{amssymb}
\usepackage{boldline,multirow,xcolor,colortbl}
\usepackage{float}

\newcommand{\ii}{\mathrm{i}}
\newcommand{\ket}[1]{\vert #1 \rangle}

\newcommand{\abs}[1]{\left| #1\right|}

\newcommand{\Real}[1]{\mathbb{R}\text{e}\left[ #1 \right]}
\newcommand{\Imag}[1]{\mathbb{I}\text{m}\left[ #1 \right]}

\newcommand{\Eq}[1]{Eq.~\eqref{#1}}

\newcommand{\Fig}[1]{Fig.~\ref{#1}}

\begin{document}

\title{Signatures of Complex Optical Response in Casimir Interactions of Type I and II Weyl Semimetals}

\author{Pablo Rodriguez-Lopez}
\affiliation{\'{A}rea de Electromagnetismo, Universidad Rey Juan Carlos,Tulip\'{a}n s/n, 28933 M\'{o}stoles, Madrid, Spain.}
\affiliation{Materials Science Factory, Instituto de Ciencia de Materiales de Madrid (ICMM),
Consejo Superior de Investigaciones Científicas (CSIC),
Sor Juana Inés de la Cruz 3, 28049 Madrid Spain.}
\affiliation{GISC-Grupo Interdisciplinar de Sistemas Complejos, 28040 Madrid, Spain.}

\author{Adrian Popescu}
\affiliation{Department of Physics, University of South Florida, Tampa, Florida 33620, USA.}

\author{Ignat Fialkovsky}
\affiliation{Centro de Matem\'atica, Computa\c{c}\~ao e Cogni\c{c}\~ao, Universidade Federal do ABC, 09210-170 Santo Andr\'e, SP, Brazil.}

\author{Nail Khusnutdinov}
\affiliation{Centro de Matem\'atica, Computa\c{c}\~ao e Cogni\c{c}\~ao, Universidade Federal do ABC, 09210-170 Santo Andr\'e, SP, Brazil.}
\affiliation{Institute of Physics, Kazan Federal University, Kremlevskaya 18, Kazan, 420008, Russia.}

\author{Lilia M. Woods$^{*,}$}
\affiliation{Department of Physics, University of South Florida, Tampa, Florida 33620, USA.}

\date{\today}

\begin{abstract}
The Casimir interaction is induced by electromagnetic fluctuations  between objects and it is strongly dependent upon the electronic and optical properties of the materials making up the objects. Here we investigate this ubiquitous interaction between Weyl semimetals, a class of 3D systems with low energy linear dispersion and nontrivial topology due to symmetry conditions and stemming from separated energy cones. A comprehensive examination of all components of the bulk conductivity tensor as well as the surface conductivity due to the Fermi arc states in real and imaginary frequency domains is presented using the Kubo formalism for Weyl semimetals with different degree of tilting of their linear energy cones. The Casimir energy is calculated using a generalized Lifhsitz approach, for which electromagnetic boundary conditions for anisotropic materials were derived and used. We find that the Casimir interaction between Weyl semimetals is metallic-like and its magnitude and characteristic distance dependence can be modified by the degree of tilting and chemical potential. The nontrivial topology plays a secondary role in the Casimir interaction of these 3D materials and thermal fluctuations are expected to have similar effects as in metallic systems.
\end{abstract}


\maketitle

\section{Introduction}

Light-matter interactions are at the forefront of current fundamental and applied research. The exchange of zero-point energy modes of fluctuating electromagnetic fields between objects with finite boundaries gives rise to the ubiquitous Casimir interaction \cite{Casimir-48}. This type of force is strongly dependent on the geometry of the objects as well as their electronic and optical response \cite{Klimchitskaya-09}. The Casimir force is especially pronounced at micro and submicrometer separations and it can limit the operation of nano and micro electronic devices due to unwanted sticktion and adhesion \cite{Chan-01}. Recent developments have shown that it is beneficial to explore the materials aspect of Casimir interactions in order to probe novel physics \cite{Woods-16}. This in turn can be of great importance to find ways to control the magnitude and sign of this ubiquitous force, which may be used to improve the performance of tiny devices. 

Materials with Dirac spectrum have been a fruitful platform for Casimir force discoveries. Perhaps the most studied system in this regard has been graphene whose reduced dimensionality and linear energy band structure led to many unusual functionalities in this interaction, including much enhanced thermal effects \cite{Bordag-06,Gomez-Santos-09,Drosdoff-10,Sarabadani-11,Svetovoy-11,Banishev-13,Khusnutdinov-18,Fialkovsky-2011}. Additionally, Casimir force phase transitions driven by external fields have been predicted in related 2D materials, such as silicene, germanene, and stanene \cite{Rodriguez-Lopez-17}. Moreover, repulsive and force quantization effects have been described in other materials, such as 3D topological insulators and Chern insulators \cite{Grushin-11,Rodriguez-Lopez-14,Wilson-15,Fialkovsky-18}. 

\begin{figure*}
\includegraphics[width=12.cm,height=4cm]{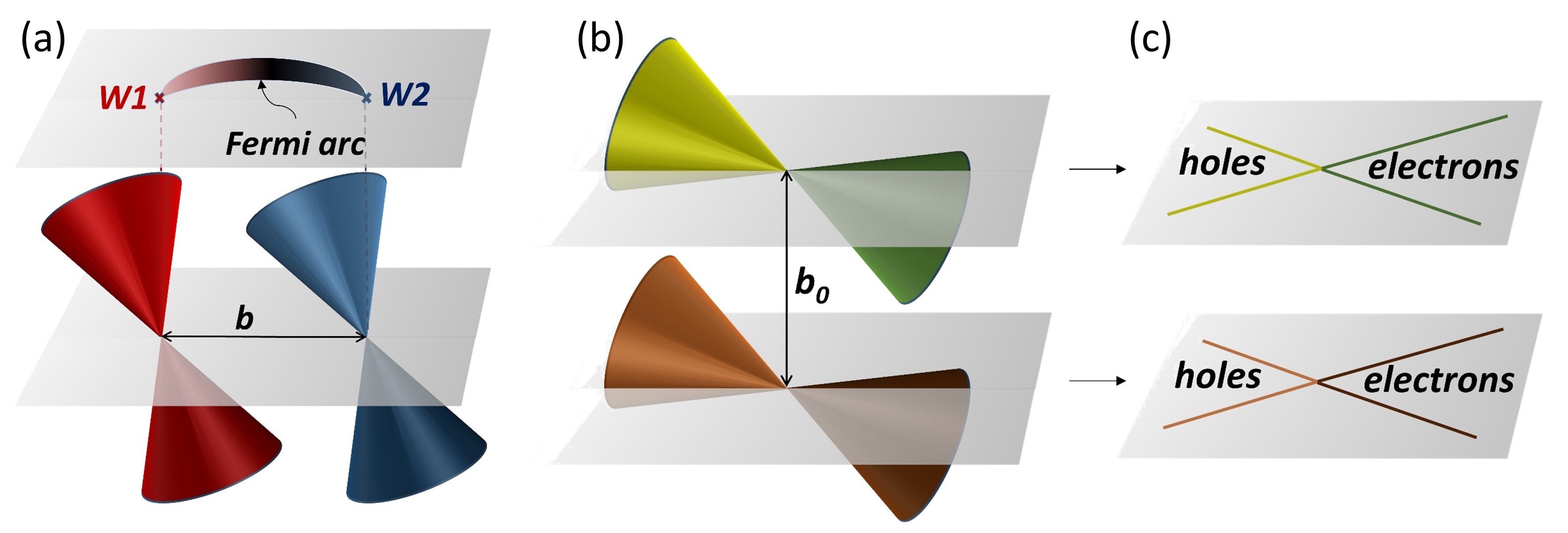}
\caption{\label{figure1} (Color online) Graphical representation of the electronic structure of Weyl semimetals. (a) Separated cones in momentum space by $b$ with tilting $\Psi<1$. The projected on the surface of the crystal bulk Weyl nodes lead to Fermi arc surface states; (b) Separated cones in energy space by $b_0$ with tilting $\Psi>1$;  (c) The severe tilting of the cones results in electron and hole states residing on the Fermi surface.}
\end{figure*}

Weyl semimetals (WSMs) are 3D materials and they have analogous to graphene linear energy dispersion band structure near the Fermi level. WSMs are also characterized by linear energy cones, whose nodes can be viewed as magnetic monopoles in reciprocal space \cite{Hasan-17}. Since these features are associated with Berry curvature, the Hall effect and its topologically nontrivial nature are also of interest \cite{Wan-2011}. Tilting of the Weyl cones is also possible as one distinguishes between type I and type II WSMs depending on the degree of tilting. It has been shown that TaAs, TaP, NbAs, and NbP (the TAAS family), Na$_3$Sb, Na$_3$Bi and some Heusler alloys are type I WSM, while MoTe$_2$, WTe$_2$, LaAlGe, and TaIrTe$_2$ belong to the type II class \cite{Armitage-2018,Yan-2016}. Characterization techniques, such as ARPES, STM, magneto-optical, and quantum oscillations measurements have been utilized to study band structure effects and optical conductivity signatures in various WSMs from both types \cite{Armitage-2018,Yan-2016}. Exotic surface states, which are associated with the nontrivial topology in these materials have also been demonstrated \cite{Wan-2011}.

All of these unique electronic structure properties have a direct consequence in the optical response of Weyl semimetals. The interplay between the 3D linear energy dispersion, separations of the Weyl cones, and their tilting results in anisotropic optical conductivity tensor with distinct features in the longitudinal and Hall components as well as the existence of surface conductivity due to the special surface Fermi arc states. In this work, using an effective model for the linear band structure and the Kubo formalism a comprehensive examination of the different optical conductivity components is presented for type I and II WSMs. The Casimir interaction is calculated using a generalized Lifshitz approach for which the boundary conditions for anisotropic media are resolved via the Fresnel reflection matrix. Although the optical response is complex, the Casimir interaction is found to be dominated by the bulk diagonal conductivity components, while the surface states and Hall conductivity play a secondary role. Unlike 2D Dirac materials for which several unusual functionalities were found due to their non-trivial, the Casimir interaction in WSMs is similar to the one for metallic systems. We find that the magnitude of the interaction is strongly by affected the cutoff energy for the linear band dispersion as well as chemical potential and cone tilting. 

\section{Properties of Weyl Semimetals}

{\it Electronic Properties} The essential electronic properties of Weyl materials with linearly dispersing bands crossing at the Fermi level can be captured by the following Hamiltonian per Weyl node \cite{Wan-2011,Soluyanov-2015}
\begin{equation}\label{eq1}
H_{\eta} = \left( \eta b_{0} + \mathbf{v}_{t}\cdot \mathbf{k} \right) \mathbf{1} + \hbar v_{F} \bm{\sigma}\cdot(\eta\mathbf{k} + \mathbf{b}), 
\end{equation}
with eigenenergies
\begin{equation}\label{eq2}
E_\mathbf{k}^{\eta} = \left( \eta b_{0} + \mathbf{v}_{t}\cdot \mathbf{k} \right) \pm \hbar v_{F}\abs{\eta  \mathbf{k} + \mathbf{b}}.
\end{equation}
This is a minimal model with a pair of Weyl nodes specified by their chirality $\eta=\pm 1$. Here $v_F$ is the Fermi velocity, $\bm{\sigma}=(\sigma_x, \sigma_y, \sigma_z)$ are the Pauli matrices, $\mathbf{k}$ is the 3D wave vector, and $\mathbf{1}$ is the $2\times 2$ identity matrix. The nontrivial topology corresponding to the above Hamiltonian is determined by the separation between the two crossing points in the band structure. For systems with broken Inversion symmetry this separation, denoted as $b_0$, marks the distance between the Weyl points in energy space, while for systems with broken Time Reversal symmetry $\mathbf{b}$ gives this separation in momentum space. These two cases are generalized in Eq. \ref{eq1} by using the four-vector $b_\mu=(b_0, \mathbf{b})$. 

In addition to their separation, the Weyl cones can also be tilted, which is determined by the tilt velocity $\mathbf{v}_t$. One distinguishes between $v_t/v_F<1$ and $v_t/v_F>1$ cases corresponding to type I and type II Weyl semimetals, respectively. The $\mathbf{v}_t=0$ situation corresponds to the simplest type I material with cones in an upright position (no tilting).  In type II WSM the tilt is large enough so the cones are tipped over transforming the point-like Fermi surface, typical for type I WSM, to a different shape. Due to the $v_t/v_F>1$ condition, there is also a finite density of states at the Fermi level characterized by electron and hole pockets. Let us note that the nontrivial topology is irrelevant of the cone tilting and it is present in both types of semimetals. Nevertheless, many properties are expected to be affected by $v_t$ and its relative value with respect to $v_F$, giving rise to distinct features in the optical response, anisotropic chiral anomaly, and anomalous Hall effects among others \cite{Yu-2016,Soluyanov-2015,Zyuzin-2016,Carbotte-2016,Mukherjee-2018}. 

In this work we consider the separation between the cones as described by $b_\mu=(b_0,b\hat{\bm{k}}_z)$ and we choose the cone tilting with respect to the $z$-axis in momentum space with  $\mathbf{v}_t=v_t\hat{\bm{k}}_z$. As a result, the above Hamiltonian and its eigenergies and eigenspinors can be simplified to 
\begin{eqnarray}
H_\eta &=& \left( b_0 + \Psi K_z^\eta\right) \mathbf{1} + \bm{\sigma}\cdot \bm{K}^\eta \label{eq3}, \\
E_\mathbf{k}^{\eta,\pm} &=& \left( \eta b_0 + \Psi\hbar v_{F}k_{z} \right)\pm |\bm{K}^\eta| \label{eq4}, \\
\ket{u^\eta_{\bm{k}}} & = & \frac{1}{\sqrt{2|\bm{K}^\eta| |\bm{M}^\eta|}}\left(\begin{array}{c}
\frac{|\bm{M}^\eta|\left( K_{x}^\eta + \ii K_{y}^\eta \right)}{K^\eta_\perp}\\
\hbar v_{F}K_\perp^\eta
\end{array}
\right), \label{eq5}
\end{eqnarray}
where $\bm{K}^\eta = \hbar v_{F}( \eta \bm{k} + b\hat{\bm{k}}_z )$, $K^\eta_\perp=\sqrt{K_x^{\eta,2}+K_y^{\eta,2}}$, and $|\bm{M}^\eta|= |\bm{K}^\eta|+K_z^\eta $. Also, $\Psi=v_t/v_F$ denotes the tilting ratio. In Fig. \ref{figure1}a, we show representative band structures for  type I WSM whose cones are separated in momentum space, which corresponds to materials with broken Time Reversal symmetry. Fig. \ref{figure1}b shows type II WSM with cones separated  along the energy direction, which reflects systems with broken Inversion symmetry. Due to the tilted cones in type II WSM free carriers at the Fermi level are also available as schematically shown in Fig. \ref{figure1}c. 

\begin{figure*}
\includegraphics[width=16.5cm,height=8.5cm]{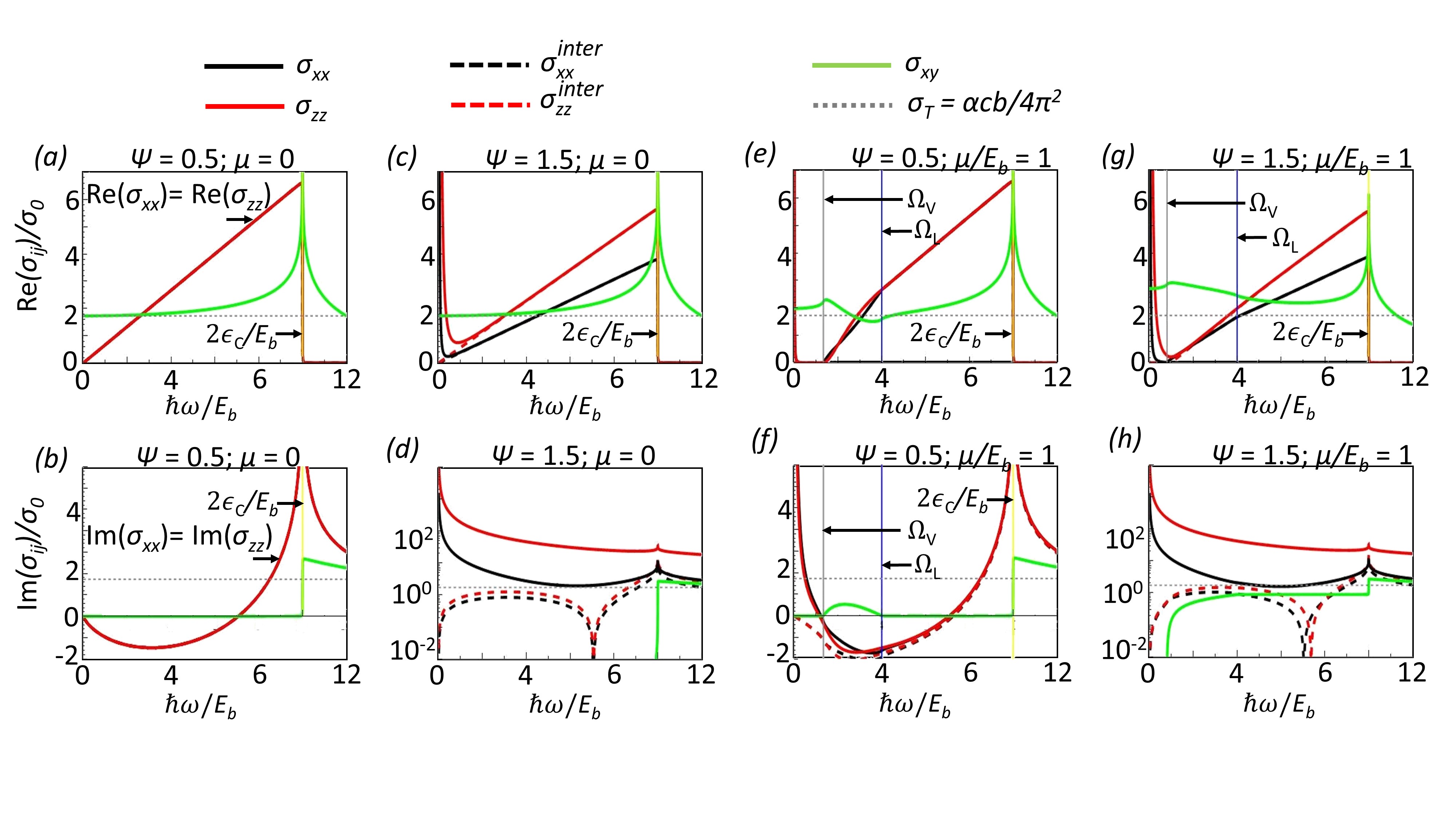}
\caption{\label{figureS1} (Color online) Real and Imaginary parts of different conductivity tensor components for a single Weyl cone scaled by $\sigma_{0} = \frac{\alpha c}{16\pi v_{F}}$ as a function of frequency scaled by $E_b=\hbar v_F b$ for: (a) and (b) $\Psi=0.5$, $\mu=0$; (c) and (d) $\Psi=1.5$, $\mu=0$; (e) and (f) $\Psi=0.5$, $\mu/E_b=1$; (g) and (h) $\Psi=1.5$, $\mu/E_b=1$. The legend given above the panels denotes the total $\sigma_{xx,b}$ (full black) and $\sigma_{zz,b}$ (full red), interband contributions $\sigma_{xx,b}^{inter}$ (dashed black) and  $\sigma_{xx,b}^{inter}$ (dashed red), the bulk Hall conductivity $\sigma_{xy,b}$ (full green) and the topological Hall conductivity $\sigma_{T}$ (dashed gray). The following parameters were used in the calculations: $\epsilon_c=10$ eV, $E_b=1$ eV, $c/v_F=300$, $\hbar/(\tau E_b)=10^{-3}$.
}
\end{figure*}

{\it Optical Response} The optical response of a given material is directly related to its electronic structure. The dynamical optical conductivity tensor is a fundamental property, which is a necessary ingredient in understanding light-matter interactions, such as the Casimir force. The components of this tensor can be calculated using the Kubo formalism \cite{mahan-book}. For a given cone specified by its chirality $\eta$, these are 
\begin{widetext}
\begin{eqnarray}\label{Kubo_Formula}
\sigma_{ij}^\eta(\omega,\textbf{q}) & = & - i e^2 \sum_{\lambda,\lambda'}\int \frac{d^{d}\textbf{k}}{(2\pi)^{d}}
\frac{<u_{\textbf{k}}^{\lambda,\eta}| v_{i}^\eta |u_{\textbf{k}+\textbf{q}}^{\lambda',\eta}> <u_{\textbf{k}+\textbf{q}}^{\lambda',\eta}| v_{j}^\eta |u_{\textbf{k}}^{\lambda,\eta}> }{\hbar(\omega + i/\tau) + E_{\textbf{k}}^{\lambda,\eta} - E_{\textbf{k}+\textbf{q}}^{\lambda',\eta} }
\frac{ n_{F}(E_{\textbf{k}}^{\lambda,\eta}) - n_{F}(E_{\textbf{k}+\textbf{q}}^{\lambda',\eta}) }{E_{\textbf{k}}^{\lambda,\eta} - E_{\textbf{k}+\textbf{q}}^{\lambda',\eta} }.
\end{eqnarray}
\end{widetext}
Here $|u_{\textbf{k}}^{\lambda,\eta}>$ and $E_{\textbf{k}}^{\lambda,\eta}$ ($\lambda, \lambda'=\pm$ correspond to the electron and hole states) are the eigenspinors and eigenenergies for the Hamiltonian in Eqs. (\ref{eq3}-\ref{eq5}) and $\textbf{v}^\eta  = \nabla_{\bm{k}}H_{s}^{\eta}/\hbar = \eta v_{F}(\sigma_{x}, \sigma_{y}, \sigma_{z})$ is the velocity operator. Also, the Fermi-Dirac distribution function is $n_F(E_{\textbf{k}}^{\lambda,\eta})=1/(e^{(E_{\bm{k}}^\lambda-\mu)/k_BT}+1)$ and $\tau$ is the relaxation time.

Using the eigenstates and energies from Eqs. (\ref{eq4}-\ref{eq5}) the conductivity tensor for both types of WSMs is found of the following form
\[
 {\bf \sigma}^\eta=
          \begin{bmatrix}
           \sigma_{xx,b}^\eta &&    \sigma_{xy,b}^\eta  &&    0  \\
            -\sigma_{xy,b}^{\eta *}  &&    \sigma_{xx,b}^\eta &&    0  \\
           0  &&    0  &&    \sigma_{zz,b}^\eta
           \end{bmatrix}.
 \]
Each bulk component is decomposed into interband  and intraband terms, such that $\sigma_{ij,b}^\eta=\sigma_{ij,b}^{inter,\eta}+\sigma_{ij,b}^{intra,\eta}$. To obtain the total conductivity corresponding to the Hamiltonian in Eq. (\ref{eq3}) summation of the contributions from the Weyl cones must be done, such that $\sigma=\sigma^{(+)}+\sigma^{(-)}$. Detailed calculations and explicit expressions for zero temperature, $T=0$, and no spatial dispersion, $\textbf{q}=0$, for the Real and Imaginary parts of the interband and intraband conductivity components are given in the Supplementary Material. 

Our results indicate that both types of materials have anisotropic optical properties with distinct diagonal components. Eqs. (\ref{eq3}-\ref{eq5}) correspond to the 3D bulk band structure, which is responsible for $\sigma_{zz,b}\neq \sigma_{xx,b}=\sigma_{yy,b}$. In addition, the nontrivial topology of the separated Weyl cones results in a surface conductivity (discussed later), which gives rise to an anisotropic surface conductivity tensor. The bulk components have an explicit dependence upon the cone tilting and the cutoff energy $\epsilon_c$, as found by several other authors \cite{Tabert-2016,Carbotte-16,Mukherjee-17,Mukherjee-18,Zyuzin-16}. The cutoff energy serves as a limit beyond which the Dirac bands are not of linear dispersion anymore \cite{Rosenstein13,Kotov16}.  This is different than the situation in graphene materials, where the optical response is found to be independent of such a cutoff \cite{Rodriguez-Lopez-18}. The unique role of $\epsilon_c$ is thus related to the interplay between the linear band dispersion and dimensionality of the material. 

The general expressions for the different conductivity components are quite complicated, as evident in the Supplementary Material. Thus in Fig. \ref{figureS1} we show the nonzero components for a single cone for both types of WSMs and their dependence on frequency. For $\omega\rightarrow 0$ $\Real \sigma_{xx,b}$ and $\Real \sigma_{zz,b}$ of type I WSM tend to zero when $\mu=0$ (Fig. \ref{figureS1}a), while these components exhibit Drude-like behavior for type II WSM due to the finite carrier concentration originating from the severe tilting (Fig. \ref{figureS1}c). The Drude-like behavior is also seen in Fig. \ref{figureS1}(e,g), in which intraband contributions from the non-zero chemical potential are included. We find that $\Real \sigma_{xx,b}$ and $\Real \sigma_{zz,b}$ exhibit a linear dependence with frequency until $2\epsilon_c$ , however there is some nonlinearity in the $(\Omega_L=\frac{2\mu}{|\Psi-1|},\Omega_U=\frac{2\mu}{|\Psi+1|})$ region in Fig. \ref{figureS1}(e,g) \cite{Carbotte-16}. The real part of the Hall conductivity shows a discontinuous behavior as $\hbar \omega$ approaches $\epsilon_c$. The imaginary parts of the different components are also given as a function of frequency. The $\Imag \sigma_{xx,b}$ and $\Imag \sigma_{zz,b}$ are discontinuous as $\hbar\omega$ approaches $\epsilon_c$ (Fig. \ref{figureS1}b,d,f,h). At small frequency the imaginary parts are mainly determined by the Drude-like behavior \cite{Tabert-2016,Mukherjee-17,Mukherjee-18}. 

In addition to the graphical representation of the WSMs optical response in Fig. \ref{figureS1}, analytical expressions in different limits for the plasma frequencies and Hall response can be found. Specifically, in both cases the intraband conductivities are of Drude type with plasma frequencies $\omega_{P}^{I,II}$ explicitly dependent on the WSM type (Eqs. (41,42,46,47) in the Supplementary Material). In addition, the plasma frequencies for WSM type II are found to diverge as a function of the cutoff, consistent with other works \cite{Tabert-2016,Mukherjee-18}. For example, in the limit of large $\epsilon_c$ the xx-component is obtained as
\begin{eqnarray}\label{eq7}
\omega_{P,xx}^{II} \approx \frac{\alpha c}{8\pi^{2}\hbar v_{F}}\left[ \left( \frac{\Psi^2 - 1}{\Psi} \right)\epsilon_{c}^{2} - 2\frac{\tilde{\mu}^{2}}{\Psi^{3}}\log(\epsilon_{c})  \right], 
\end{eqnarray}
where $\tilde{\mu}=|\mu-b_0|$. 
Similar second power law and logarithimic divergencies upon $\epsilon_c$ are present for the zz-component of the bulk intraband conductivity. 

\begin{figure*}
\includegraphics[width=13.5cm,height=7cm]{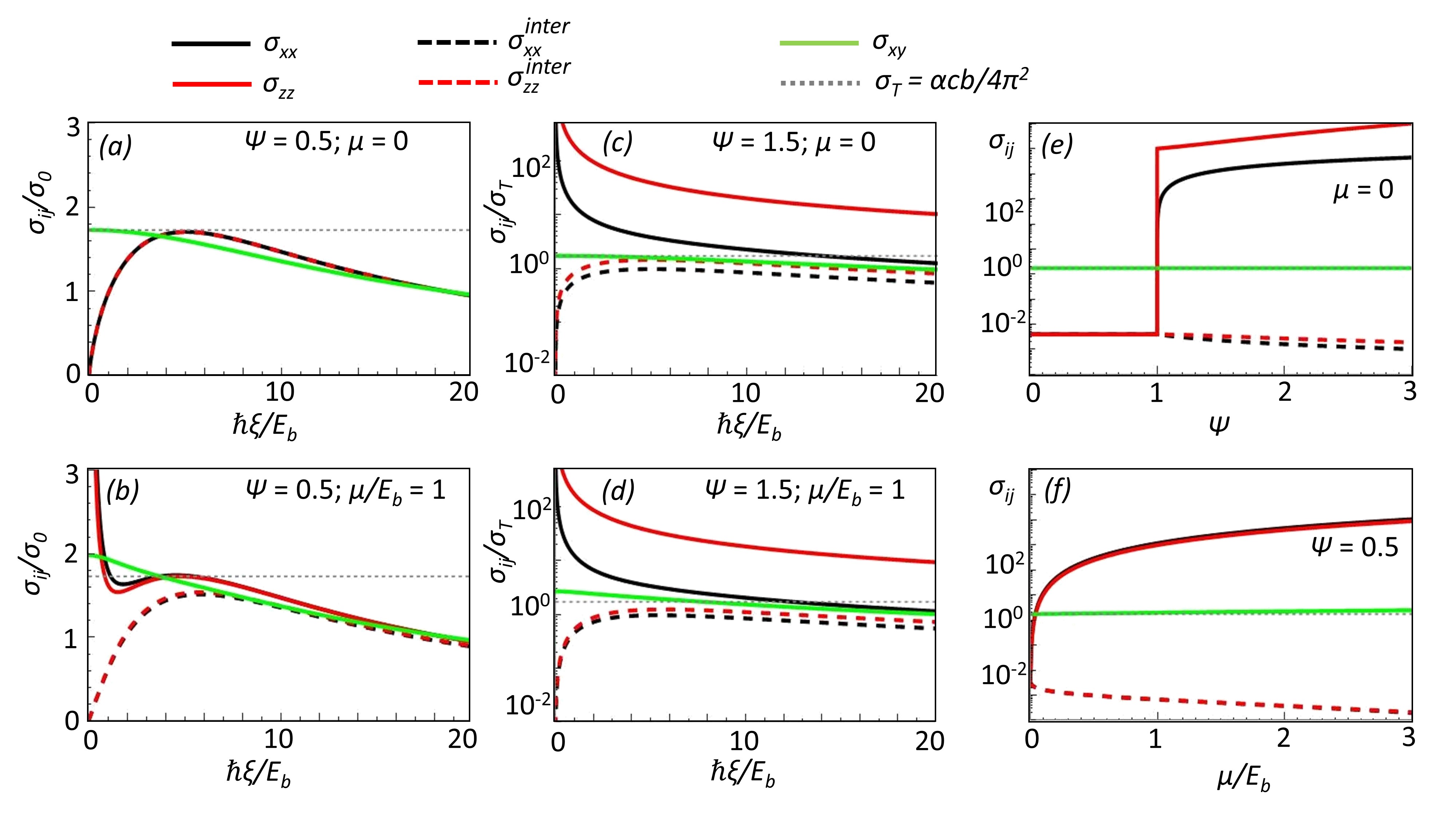}
\caption{\label{figure3} (Color online) Different contributions to the bulk conductivity components $\sigma_{ij}$ scaled by $\sigma_{0} = \frac{\alpha \hbar c}{16\pi\hbar v_{F}}$ as a function of imaginary frequency $\xi$ scaled by $E_b=\hbar v_F b$ for: (a) $\Psi=0.5$, $\mu=0$; (b) $\Psi=0.5$, $\mu/E_b=1$; (c) $\Psi=1.5$, $\mu=0$; (d) $\Psi=1.5$, $\mu/E_b=1$. Different contributions to the bulk conductivity components $\sigma_{ij,b}$ scaled by $\sigma_{0}$ at $\xi=0$ as a function of: (e) $\Psi$ for $\mu=0$ and (f) $\mu/E_b$ for $\Psi=0.5$. The following parameters were used in the calculations: $\epsilon_c=10$ eV, $E_b=1$ eV, $c/v_F=300$, $\hbar/(\tau E_b)=10^{-3}$.}
\end{figure*}

It is also interesting to examine the behavior of the plasma frequencies as a function of chemical potential. For type I WSM $\lim_{\tilde{\mu} \to 0}\omega_{P,xx(zz)}^I=0$, which shows vanishing intraband conductivity as the chemical potential passes through the Fermi level. For type II WSM, one finds that 
\begin{equation}\label{eq8}
\lim_{\tilde{\mu} \to 0}\omega_{P,xx}^{II} = \frac{\alpha c}{8\pi^{2}\hbar v_{F}}\frac{\Psi^{2} - 1}{\Psi}\left( \epsilon_{c}^{2} + (\hbar v_F b)^2 \right).
\end{equation}
Thus even if $\tilde{\mu} = 0$, there is a nonzero intraband conductivity. This correlates with the severe tilting in the WSM from $\Psi>1$ which results in a band crossing at the Fermi level, meaning that there are always free carriers available in the system (Fig. \ref{figure1}c). 

We further study the Hall conductivity for each type of WSM. The zero frequency limit, which is a measure of the anomalous Hall effect, is obtained as 
\begin{eqnarray}
\sigma^{I,\eta}_{xy}(\omega=0) & = & \frac{\alpha c}{4\pi^{2}}\left[ \hbar b + \frac{\eta\tilde{\mu}}{\hbar v_{F}}\frac{\tanh ^{-1}(\Psi )-\Psi }{\Psi ^2} \right], \label{eq9} \\
\sigma_{xy}^{II,\eta}(\omega=0) &=& \frac{\alpha c}{4\pi^{2}} \left[ \hbar b  
+\frac{\eta\tilde{\mu}}{\hbar v_{F}\Psi^{2}}\log \frac{\epsilon_c\Psi\sqrt{ \Psi^{2} - 1 }}{\tilde{\mu}} \right]. \label{eq10}
\end{eqnarray}
The first term in the above expressions is common to both types of WSMs. It is proportional to the distance $b$ between the cones in momentum space and it reflects the nontrivial topology of the materials. The second term in each $\sigma_{xy}^{I(II),\eta}$ comes from the tilting of the cones and the chemical potential scaled by the separation in energy space $\tilde{\mu}=|\mu-b_0|$ and it does not have topological origin. The Hall conductivity of type II WSM also shows a logarithmic divergence with the energy cutoff, while $\sigma_{xy}^{I,\eta}$ is finite at the large $\epsilon_c$ limit. Our results are consistent with calculations reported by other authors \cite{Mukherjee-17,Zyuzin-16}. Let us note, however that the total Hall conductivity for both types of Weyl semimetals is $\sigma^{I(II)}_{xy}(\omega=0)=\frac{\alpha c \hbar b}{4\pi^2}$ as the second terms in Eqs. (\ref{eq9}, \ref{eq10}) cancel out upon addition due to the opposite in sign cone chiralities.  

{\it Surface Conductivity due to Fermi Arcs} One of the hallmarks of WSMs is their Fermi arc surface states. Other materials support Fermi surface states that have circular or deformed circular shapes \cite{Inglesfield-82,Wang-13}. In a Weyl semimetal occupying semi-infinite space with an abrupt boundary, however, only half of this usual shape resides on the surface giving rise to a Fermi arc, which essentially starts and ends at the surface projections of the bulk Weyl points, as schematically shown in Fig. \ref{figure1}a. The Fermi arcs are also a signature of the nontrivial bulk topology of WSMs regardless of tilting and their existence is determined by the bulk-surface correspondence \cite{Xu-15,McCormick-2017,Mathai-17,Mong-2011,Buividovich-14}. In the case of Weyl cones separated in momentum space the Fermi arc can be represented as a single Dirac fermion (similar to graphene), whose conductivity in the low frequency is of Drude-like type and it is determined by the distance between the cones \cite{Grushin-15}. Specifically, for the case of  $b_\mu=(0,b \hat{\bf k}_z)$ at the $x=0$ surface one finds there is only one nonzero conductivity component
\begin{equation} \label{eq11}
\sigma_{yy,s} = \frac{-i\alpha c v_F b}{2\pi^2(\omega+i\Gamma)}.  
\end{equation}
At the $y=0$ boundary, the nonzero surface conductivity is $\sigma_{xx,s}$ with the same value as the above equation. There is no surface conductivity at the $z=0$ boundary since the projection of the vector $b_\mu=(0,b\hat{\bm{k}}_z)$ is zero in this case \cite{Grushin-15}. Thus this optical response is associated with the WSM boundary due to the distance between the cones $b$ and it is highly anisotropic. 

{\it Optical Response at Imaginary Frequency} For the Casimir interaction the conductivity components in imaginary frequency domain $\omega=i\xi$ are needed. Thus it is convenient to use the Kramers-Kr\"onig relations for which,
\begin{eqnarray}\label{KKTransform}
\Real{\sigma_{ij}(\ii\xi)} & = & \frac{2}{\pi}\int_{0}^{\infty}d\omega\frac{\xi}{\omega^{2} + \xi^{2}}\Real{\sigma_{ij}(\omega)}\nonumber\\
& = & \frac{2}{\pi}\int_{0}^{\infty}d\omega\frac{\omega}{\omega^{2} + \xi^{2}}\Imag{\sigma_{ij}(\omega)}.
\end{eqnarray}
Realizing that $\Imag{\sigma_{ij}(\ii\xi)} = 0$, the analytical continuation of  $\Real{\sigma_{ij}(\ii\xi)}$ to all positive imaginary frequencies in the upper half of the complex plane gives the final expressions for the conductivity components. By taking $\xi \to\xi + \hbar/\tau $ finite dissipation can be included in the optical response.

In \Fig{figure3} results for some functional dependences of the conductivity components, commensurate with the ones in \Fig{figureS1} , for a single cone at imaginary frequency $\omega=i\xi$ are given. In all cases $\sigma_{xx,b}^{inter}$ and $\sigma_{zz,b}^{inter}$ increase linearly in the small $\xi$ region (\Fig{figure3} a,b,c,d). The Hall conductivity has a very weak dependence upon $\xi$ and in both WSMs the $\xi=0$ limit is governed by the analytical expressions in \Eq{eq7} and \Eq{eq8}). There is a discontinuous jump at $\Psi=1$, which marks the transition between type I and type II WSMs as shown in \Fig{figure3}e.  As $\Psi$ increases for type II WSM the conductivity is determined primarily by the intraband contributions due to increasing the number of free carriers from the severe tilting even if $\mu=0$. Also, $\sigma_{ij,b}$ as a function of $\mu$ is depicted for type I WSM in \Fig{figure3}e at $\xi=0$. This behavior is consistent with our analytical considerations and it shows that the response is mainly determined by the intraband contributions as the chemical potential grows.    

\section{Characteristic Behavior of the Casimir Interaction}

The Casimir interaction can be calculated using the Lifshitz approach \cite{Klimchitskaya-09,Woods-16}, however for lossy materials with reflection matrices with complex frequency dependence this formalism must be applied carefully. Specifically, one must account for contributions from propagating and evanescent waves, which can arise from frequencies above and below the light cone, respectively \cite{Genet2003}. Thus, for semi-infinite planar objects taken to be separated by a distance $d$ along the $z$-axis, the interaction energy per unit area $A$ can be written as 
\begin{eqnarray}\label{Lifshitz_Formula}
\frac{E}{A} &=& \frac{\mathcal{E}}{A} + \frac{\mathcal{E}^*}{A}, \\
\frac{\mathcal{E}}{A} &=& k_{\text{B}}T {\sum_{n}}' \int\frac{d^{2}{\bf k}_{\|}}{(2\pi)^2}\log \det \left( {1 - {\bf R}_{1}\cdot {\bf R}'_{2} e^{-2 k_{n} d}}\right).
\end{eqnarray}
Here ${\bf R}_{j} = {\bf R}_{j} (k_n, k_x, k_y)$ for $j=1,2$ are the $2\times2$ Fresnel reflection matrices for the interacting objects, where $k_{n}=\sqrt{k_x^2+k_y^2+\xi_n^2/c^2}$ with $\xi_n=2 \pi n k_{\text{B}}T /\hbar$ being the Matsubara frequencies ($n=0,1,2,\ldots$ and ${\bf k}_{\|}=(k_x, k_y)$). Also, 
${\bf R}'_{j} = {\bf R}_{j} (-k_n, k_x, k_y)$ denotes the reflection matrix for scattering with opposite direction \cite{Fialkovsky-18}. The generalized Lifshitz expression from Eqs. \ref{Lifshitz_Formula} is necessary for the interaction involving Weyl semimetals. From our subsequent calculations we find that the reflection matrices may not be real in some imaginary frequency ranges. This is attributed to the interplay between propagating and evanescent wave stemming from the anisotropic optical response, which motivates the generalization of the Lifshitz formalism \cite{Genet2003}. 

\begin{figure}
\includegraphics[scale=0.12]{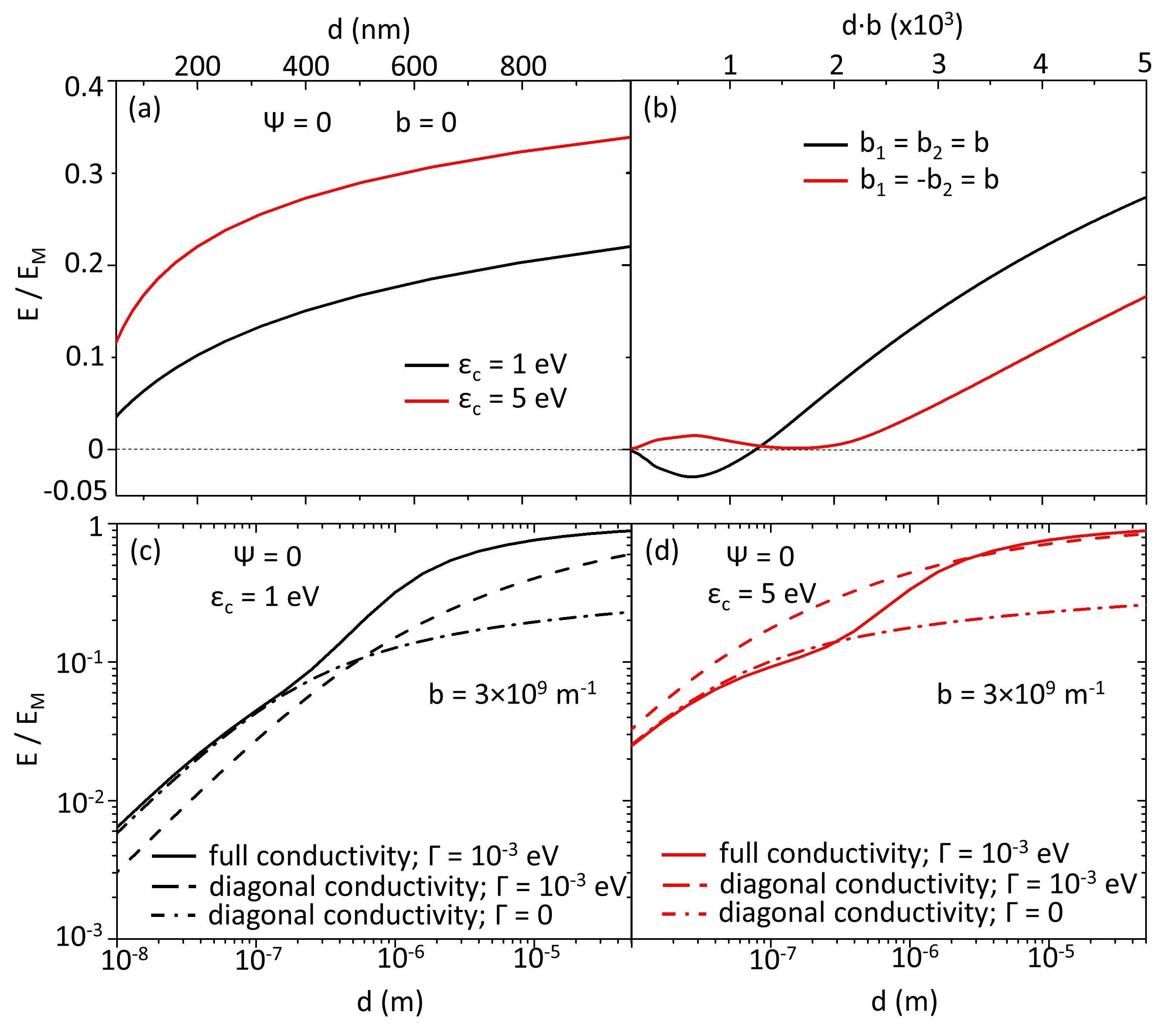}
\caption{\label{figure4} (Color online) The Casimir energy scaled by $E_M=-\frac{\pi^2\hbar c}{720 d^3}$ as a function of: (a) distance for coinciding nontilted Weyl cones ($b=0, \Psi=0$)  and different values of the cutoff parameter $\epsilon_c$; (b) the unitless parameter $d\cdot b$ with bulk $\epsilon_{xx}=\epsilon_{yy}=\epsilon_{zz}=1$ and no surface conductivity; (c) and (d) distance for separated nontilted Weyl cones with $b_1=b_2=b$ calculated with the full conductivity (diagonal and Hall components) with $\Gamma=10^{-3}$ eV (full curves), diagonal conductivity and no Hall components with $\Gamma=10^{-3}$ eV (dashed curves), and diagonal conductivity and no Hall components with $\Gamma=0$ (dotted curves) for $\epsilon_c=1$ eV and $\epsilon_c=5$ eV respectively. Also, $b_{1,2}$ denote the cone separations of the two interacting Weyl materials. 
}
\end{figure}

The reflection matrices depend explicitly on the optical response of the involved materials and they are obtained by applying electromagnetic boundary conditions for semi-infinite 3D objects with planar surfaces separated at a finite distance. Resolving the reflection matrices is technically difficult as each type of Weyl semimetal has anisotropic bulk and surface optical response as well as nonzero bulk Hall response.  Nevertheless, the electric and magnetic fields can be expanded in terms of the suitable for planar geometry orthogonal functions $\bm{M}_{\bm{k}}(\bm{r})  =  \frac{1}{k_{\parallel}}\nabla\times\left[\hat{\bm{z}}\phi_{\bm{k}}(\bm{r})\right]$ and $\bm{N}_{\bm{k}}(\bm{r})  =  \frac{c}{\omega}\nabla\times\bm{M}_{\bm{k}}(\bm{r})$ ($\phi_{\bm{k}}(\bm{r}) = e^{\ii\bm{k}\cdot\bm{r}}$, $k_{\parallel} = \sqrt{ k_{x}^{2} + k_{y}^{2} }$). Details of the derivation of the electromagnetic radiation impinging on the $x=0$, $y=0$, or $z=0$ surface of a semi-infinite Weyl material can be found in the Supplementary Material.

Many of the topologically nontrivial features in Weyl semimetals are related to the separation of their energy cones in momentum space. In order to better understand this issue in the context of the Casimir interaction, we consider some limiting cases with respect to $b_\mu=(0,b\hat{\bm{k}}_z)$. The so obtained results are then compared with the numerical calculations, in which the full optical response of the materials is taken into account.

{\it Coinciding Weyl Cones} At this point, we consider non-tilted Weyl cones with $b=0$. Such a situation with coinciding cones corresponds to a 3D Dirac semimetal \cite{Armitage-2018}.   In this case, one finds that the interband conductivity components in the imaginary frequency domain are of the following form 
\begin{eqnarray} \label{eq15}
\sigma_{ij}^{inter}(\xi) & = & \frac{\alpha c \xi}{12\pi^{2} v_{F}}\log\left(\frac{\hbar^{2}\xi^{2} + 4\epsilon_{c}^{2}}{\hbar^{2}\xi^{2} + 4\mu^{2}}\right)\delta_{ij}.
\end{eqnarray}
This isotropic response simplifies the reflection matrices entering in Eqs. (\ref{Lifshitz_Formula}) and some asymptotic behavior of the Casimir energy can be found analytically. For example,  the quantum mechanical ($T=0$) Casimir energy in the limit of large cutoff frequency is obtained as the energy for perfect metals, such that $ \frac{E_M}{A}=-\frac{\pi^2\hbar c}{720d^3}$. As $\epsilon_c$ is finite, the limiting cases of large and small separations with respect to the characteristic distance $d_0=\frac{\hbar c}{2\epsilon_c}$ can be  found
\begin{eqnarray}\label{Small_Distance_Limit_WSM}
\lim_{d\ll d_{0}} E  & = & \frac{45}{\pi^{4}}\frac{d}{d_{0}}E_{M}\\
& & \times\left[ \frac{1}{12 \beta ^2} + 1 +\beta +e^{\beta } (\beta +2) \beta  \text{Ei}(-\beta ) \right],\nonumber\\
\label{Large_Distance_Limit_WSM}
\lim_{d \gg d_{0}} E & = & \frac{45}{\pi^{4}}\frac{X^3-3X+2}{X(1+X)^2} E_M,
\end{eqnarray}
where $\beta = \frac{3\pi v_{F}}{\alpha c}$, $X=\sqrt{ 1 + \frac{1}{\beta }\log\left(1 + \frac{d^{2}}{d_{0}^{2}}\right) }$, and $\text{Ei}(x )$ is the exponetial integral evaluated at $x$. 
Thus the small $d$ limit results in a $d^{-2}$ scaling law for the energy. Due to the weak logarithmic dependence on $d$ in $X$, however, the large $d$ limit is essentially similar to the one for perfect metals with a $d^{-3}$ scaling law in the energy. 

In Fig. \ref{figure4}a we show the numerical calculations for the quantum mechanical interacting energy of materials with optical response from Eq. (\ref{eq15}). These results indicate that the energy cutoff practically determines the characteristic behavior of the Casimir energy through the distance $d_0$, which is in agreement with Eqs. (\ref{Small_Distance_Limit_WSM}), (\ref{Large_Distance_Limit_WSM}). Specifically, for $\epsilon_c=1$ eV one finds that $d_0=100$ nm, while for $\epsilon_c=5$ eV - $d_0=20$ nm. Thus the onset of a metallic-like interaction is strongly related to the range of validity of the linear band structure approximation controlled by $\epsilon_c$. Such an explicit dependence upon the cutoff energy is not present in 2D materials with Dirac energy spectra, where the optical response and Casimir energy are independent of the bandwidth of the linear bands, except at very small distances \cite{Rodriguez-Lopez-14,Rodriguez-Lopez-17,Rodriguez-Lopez-18}. 

\begin{figure*}
\includegraphics[width=18cm,height=4.5cm]{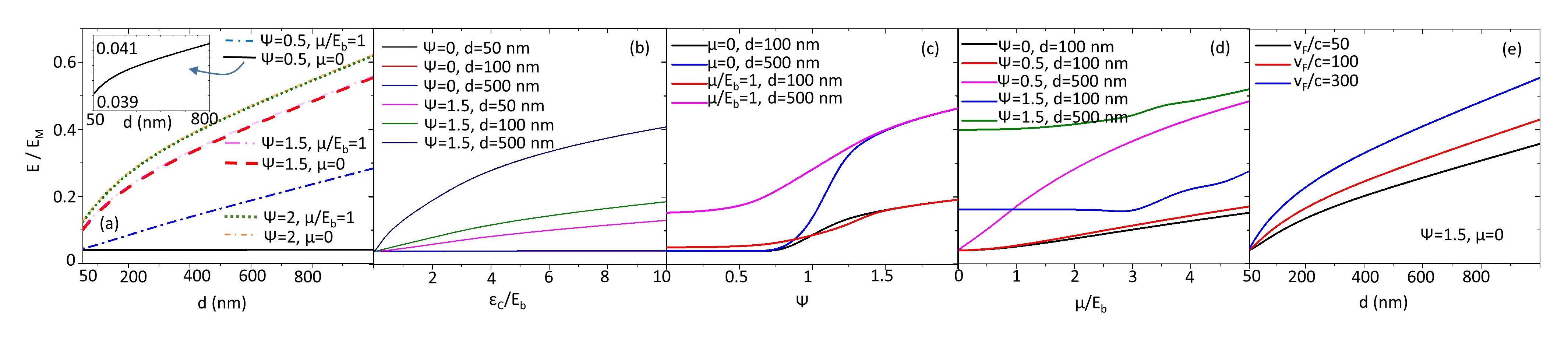}
\caption{\label{figure5} (Color online) The Casimir energy scaled by $E_M=-\frac{\pi^2\hbar c}{720 d^3}$ for two identical Weyl semimetals separated along the y-direction as a function of: a) distance for different values of cone tilting and chemical potential ($\epsilon_c=5$ eV, $E_b=1$ eV, $c/v_F=300$). The vertical axis of the inset is in log scale for better visibility; (b) cutoff energy scaled by $E_b$ for different separations ($\epsilon_c=5$ eV, $\Psi=0$, $c/v_F=300$, $\mu=0$); (c) tilting for different separations and chemical potentials ($\epsilon_c=5$ eV, $E_b=1$ eV, $c/v_F=300$, $\mu=0$); (d) chemical potential for different titling and separations ($\epsilon_c=5$ eV, $E_b=1$ eV, $c/v_F=300$); (e) distance for different values of the Fermi velocity ($\epsilon_c=5$ eV, $E_b=1$ eV, $\mu=0$, $\Psi=0.5$). In all calculations $\hbar/(\tau E_b)= 10^{-3}$. 
}
\end{figure*} 

{\it The Role of Cone Separation} The parameter $b$ enters into the reflection matrices in a complicated way, which makes it difficult to obtain analytical expressions for the Casimir energy for the considered materials. To get some idea about the role of the cone separation, we consider the situation when there is no surface conductivity and the bulk conductivity tensor contains only off-diagonal components with $\sigma_{xy,b}^\eta=\sigma_T=\frac{e^{2}b}{4\pi^2}$, which capture the anomalous Hall effect (Eqs. (\ref{eq9}, \ref{eq10})). We find that in the limiting case of small distance, the Casimir energy is
\begin{equation} \label{eq18}
\lim_{d \rightarrow 0}\frac{E}{A}  =  \frac{\hbar c\alpha^{2}}{96\pi^{4} d}b_{1} b_{2},
\end{equation}
where $b_{1,2}$ are the cone separation parameters for the two interacting materials. In the large distance limit, the energy for perfect metals is obtained. 

Eq. (\ref{eq18}) shows a $d^{-1}$ scaling law in the short distance separation and the $\alpha^{2}$ dependence in the numerator indicates that the energy is of rather small magnitude.  It is also recognized that the product $b_{1}b_{2}$ can be positive or negative given that the cone separation can be in positive or negative domains in momentum space \cite{Armitage-2018,Yan-2016}. This behavior is reminiscent of the Casimir interaction in other topological materials, whose response is dominated by the Hall conductivity  \cite{Rodriguez-Lopez-17,Grushin-11,Rodriguez-Lopez-14}. The positive sign of the product of the cone separation parameters, which are essentially proportional to the constant anomalous Hall conductivity, is associated with repulsion as found in 3D topological insulators and 2D Chern insulators. However, since $b_{1,2}$ are continuous, the Casimir energy is not quantized as found in other topological materials due to the discrete nature of their Hall conductivities. Additionally, the Casimir energy in topological and Chern insulators has $d^{-3}$ scaling law, while in Weyl semimetals there is a crossover in distance dependence from $d^{-1}$ at small separations to $d^{-3}$ at large separations.

The numerical results for the interaction energy when only the bulk Hall conductivity $\sigma_{xy,b}^\eta=\sigma_T=\frac{e^{2}b}{4\pi^2}$ is taken into account are shown in Fig. \ref{figure4}b. The repulsive interaction at small separations is dictated by the sign of the $b_1b_2$ product according to Eq. (\ref{eq18}), while at large separations the energy approaches the limit of perfect metals, which is in agreement with the above discussed analytical expressions. The strongest repulsion is found for $b\cdot d (\times 10^3)\sim 1/2$, thus characteristic values $b=5-20$ nm$^{-1}$ correspond to $d=0.1-0.025$ microns. Nevertheless, this repulsive interaction is at least two orders of magnitude smaller than $E_M$ in the sub-micron and micron separation region (depending on $b$ values), as suggested by Fig. \ref{figure4}b.  

In Fig. \ref{figure4}c,d, the numerical results for the normalized Casimir energy as a function of separation calculated by taking into account the full optical conductivity tensor of dissipative materials are shown for two values of the cutoff energy. The corresponding energies obtained when only the diagonal components of the conductivity with and without dissipation are taken into account, are  also shown. These results reveal the surprising role of dissipation: while the low and high $d$ limits in $E$ are best captured for calculations with the diagonal conductivity (no Hall components), the low $d$ limit is better described when $\Gamma=0$, while the higher $d$ range is better described when dissipation is finite. Fig. \ref{figure4}c,d further shows that the role of cone separation is to bring $E$ closer to $E_M$ in the intermediate distance range and and all repulsive effects are washed out by the dominant diagonal components of the optical response. 

At this point, one might ask how the surface conductivity due to the Fermi arcs affects the Weyl Casimir interaction. For the chosen $b_\mu=(0,b\hat{k}_z)$ 
$\sigma_{zz,s}=0$ as discussed earlier, thus to probe this property WSMs separated along the $y$- or $x$-axis must be considered with the appropriate Fresnel matrices given in the Supplementary Material. Although the numerical calculations are technically difficult we find that the surface conductivity plays even a lesser effect in the Casimir energy as compared to the one of the Hall conductivity as discussed above. The surface conductivity increases the metallic-like nature of the interaction, but given the smallness of the effect further results are not given. 

{\it Quantum Mechanical  Casimir Energy for Tilted WSMs} In what follows, we further investigate how other characteristics affect the interaction between Weyl materials. More specifically, by utilizing the full optical response dissipative type I and type II WSMs separated along the $z$-axis are considered and the Casimir energy upon cutoff energy, degree of tilting, and chemical potential is calculated. For this purpose, the generalized Lifshitz formula in Eq. (\ref{Lifshitz_Formula}), while the optical response and Fresnel matrices can be found in the Supplementary Material. These results help us find ways to modulate the Weyl Casimir interaction as well as recognize signatures of the optical response and compare with the above discussed analytical expressions. 

The scaling dependence shown in Fig. \ref{figure5}a indicates that Casimir energy increases as $d$ is increased, although for the case of $\Psi=0$, $\mu=0$ this increase is at a much smaller scale as compared to the other displayed cases. This behavior is reminiscent of the interaction between Drude-like metals controlled by the magnitude of their plasma frequencies. The larger the plasma frequencies, the closer the interaction is to the one for perfect metals. In the case of larger $\Psi$ and $\mu$, the intraband conductivities dominate the Casimir interaction due to their large $\omega_P$, while for $\Psi=0$ and $\mu=0$ the optical response is dominated by the interband components and $E/E_M$ is a small fraction.  

The behavior of $E/E_M$ vs $d$ for $\Psi=0$, $\mu=0$ in Fig. \ref{figure5}a is actually very similar to the analytical results of the Casimir energy for coinciding Weyl cones whose $d_0=20$ nm when $\epsilon_c/E_b=5$ (Fig. \ref{figure4}a). Indeed, $E/E_M$ at large distances ($d\gg d_0$) is practically determined by the prefactor in Eq. (\ref{Small_Distance_Limit_WSM}), which indicates that type I WSM with $\mu=0$ behaves as a Dirac material with coinciding cones in the displayed region of distance separation. For the other cases in Fig. 5a, however, the nonzero plasma frequency elevates the role of the intraband conductivity and the interaction is essentially similar to the one of Drude-like metals \cite{Klimchitskaya-09}. Given that the cutoff energy plays a prominent role in $\omega_{P}^{II}$ (Eqs. \ref{eq7}, \ref{eq8}), Fig. \ref{figure5}b shows how the Casimir interaction changes as a function of $\epsilon_c$ for type I and II WSMs. The much more pronounced enhancement of $E$ for type II WSMs as $\epsilon_c$ is increased is due to the large intraband optical conductivity and specifically the strong dependence of its plasma frequency on the energy cutoff (Eqs. (\ref{eq7}, \ref{eq8})). 

Fig. \ref{figure5}c displays how the Casimir energy changes as a function of degree of cone tilting. The change of an almost flat $E/E_M$ to a linear-like behavior at $\Psi\sim 1$ signals 
a transition in the Casimir energy from type I to type II WSM. Again, the much stronger interaction for type II WSMs as compared to type II WSMs in Fig. \ref{figure5}c gives another perspective of the much enhanced role of the intraband optical conductivity for $\Psi>1$. Fig. \ref{figure5}d further shows that the chemical potential is also an effective parameter in controlling the magnitude of the Casimir energy for both types of WSMs, although the degree of tilting and distance separation can affect the degree of change. Additionally, increasing the Fermi velocity of the linear bands in Eq. (\ref{eq4}) can increase the magnitude of Casimir energy although the characteristic behavior in terms of distance and other parameters are preserved (Fig. \ref{figure4}e).

Our numerical calculations indicate that the $b$ parameter plays a rather small role and the Casimir interaction is dominated by the contributions from the diagonal components in the optical tensor. This can be easily understood in the context of the above discussions for the limiting cases of coinciding Weyl cones and materials with only Hall conductivity components. In fact, taking the case of non-tilted Weyl cones and expanding the Fresnel reflection matrices for small $\sigma_{xy}$ due to $\sigma_{xy}\ll \sigma_{xx}=\sigma_{zz}$, we find that the first order correction in the small distance limit between Weyl semimetals with $\mu=0$ is the same as Eq. (\ref{eq18}). Given the much longer range due to the $d^{-1}$ dependence and the $\alpha^2$ in the numerator, it becomes clear that first order correction is much smaller than the dominant Casimir energy due to the diagonal components of the bulk conductivity response.

{\it Thermal Casimir energy}
Thermal fluctuations in the Casimir energy are captured in the $n=0$ Matsubara term of \Eq{Lifshitz_Formula} and they are expected to dominate at room temperature at separations on the sub-micron and micron scales for many materials \cite{Klimchitskaya-09,Woods-16}. To study the thermal effects in the Casimir interaction of WSMs, we calculate the conductivity at finite $T$ in the limit $\xi\to0^{+}$ using the results in the Supplementary Material (Bulk Optical Conductivity) and the Maldague's formula \cite{Rodriguez-Lopez-17,Giuliani2005,Maldague78}. The interband conductivity is linear in $\xi$ in the limit of small frequency and it is well approximated as
\begin{eqnarray} \label{eq19}
\sigma_{ij}(\xi) & \approx & \frac{2\alpha c\xi}{3\pi v_{F}}\log\left(\frac{\epsilon_{c}}{\text{Max}(|\mu|,\frac{\pi k_BT}{2e^\gamma})}  \right)\delta_{ij},
\end{eqnarray}
where $\gamma$ is the Euler's constant. The intraband conductivity at finite $T$ and $\xi\to0$ has the typical Drude-like expression
\begin{equation} \label{eq20}
\sigma_{ij}(\xi)\approx \frac{c^{2}\omega_{P}^{(I,II) 2}}{4\pi(\xi + \Gamma)}\delta_{ij},
\end{equation}
where $\omega_P^{I,II}$ are the plasma frequencies for type I and type II WSMs. Since the optical response at $\xi\to0^{+}$ is dominated by the intraband contribution, the thermal Casimir energy is found as 
\begin{eqnarray} \label{eq21}
\frac{E_{T}}{A} & = & - \frac{k_{B}T}{16\pi d^{2}}\zeta(3),
\end{eqnarray}
where $\zeta(s)$ is the Riemann zeta function. Thus the thermal effects in the WSM Casimir interaction are of the usual form of a Drude metal.

\section{Conclusions} 
The interaction induced by electromagnetic fluctuations between Weyl materials is determined by the complicated interplay between the optical and electronic response properties of these systems. In this study, we present a comprehensive investigation of the bulk and surface conductivity components by distinguishing between type I and type II WSMs. Our results indicate that the Casimir interaction exhibits a behaviors similar to the one of metallic-like systems due to the dominant role of the diagonal bulk conductivity components. The quantum mechanical interaction can be modulated as a function of chemical potential and cone tilting as captured by the $\mu$ and $\Psi$ dependences in the plasma frequency for type I and type II materials. 

The explicit $\epsilon_c$ dependence in the bulk conductivity  renders strong dependence of the interaction upon the cutoff energy corresponding to the validity of the linear energy dispersion. Unlike topological and Chern insulator, where the nontrivial topology can result in significant repulsive and even quantization effects, the anomalous Hall conductivity plays a rather small role in the quantum interaction in Weyl materials. We also find that because of the dominant diagonal bulk conductivity, thermal fluctuations are expected to affect the Casimir interaction in a similar way as the case of metallic systems. Our investigation is a testament that a thorough understanding of the fundamental electronic and optical properties of materials is necessary in order to make progress towards other research areas, such as light-matter interactions and the Casimir force as a particular example.

\begin{acknowledgments}
P.R.-L. acknowledges partial support from TerMic (Grant No. FIS2014-52486-R, Spanish Government), CONTRACT (Grant No. FIS2017-83709-R, Spanish Government, from Juan de la Cierva - Incorporacion program (Ref: I JCI-2015-25315, Spanish Government), MAT2015-66888-C3-1R and RTI2018-097895-B-C41 funded by the Spanish Government (Ministerio de Ciencia, Innovación y Universidades) and by the Feder Program of the European Union. 
N.K. was supported in parts by the grants 2016/03319-6, 2017/50294-1 and 2019/10719-9 of the São Paulo Research Foundation (FAPESP), by the RFBR projects 19-02-00496-a. 
L.M.W. acknowledges funcial support from the US Department of Energy under grant No. DE-FG02-06ER46297. 
Helpful discussions with Drs. Pablo San-Jose, Alberto Cortijo and Adolfo G. Grushin are also acknowledged. 
\end{acknowledgments}

\vspace{0.2cm}

{\bf Conflict of Interests:} The authors declare no conflict of interest.


\end{document}


\title{Supplementary Material: Signatures of Complex Optical Response in Casimir Interactions of Type I and II Weyl Semimetals}

\author{Pablo Rodriguez-Lopez}
\affiliation{\'{A}rea de Electromagnetismo, Universidad Rey Juan Carlos,Tulip\'{a}n s/n, 28933 M\'{o}stoles, Madrid, Spain.}
\affiliation{Materials Science Factory, Instituto de Ciencia de Materiales de Madrid (ICMM),
Consejo Superior de Investigaciones Científicas (CSIC),
Sor Juana Inés de la Cruz 3, 28049 Madrid Spain.}
\affiliation{GISC-Grupo Interdisciplinar de Sistemas Complejos, 28040 Madrid, Spain.}

\author{Adrian Popescu}
\affiliation{Department of Physics, University of South Florida, Tampa, Florida 33620, USA.}

\author{Ignat Fialkovsky}
\affiliation{CMCC Universidade Federal do ABC, Avenida dos Estados 5001, CEP 09210-580, Sao Paolo, Brazil.}

\author{Nail Khusnutdinov}
\affiliation{CMCC Universidade Federal do ABC, Avenida dos Estados 5001, CEP 09210-580, Sao Paolo, Brazil.}
\affiliation{Institute of Physics, Kazan Federal University, Kremlevskaya 18, Kazan, 420008, Russia.}

\author{Lilia M. Woods}
\affiliation{Department of Physics, University of South Florida, Tampa, Florida 33620, USA.}

\date{\today}

\maketitle


\section{Bulk Optical Conductivity Tensor Components}
Here we summarize our results for the nonzero conductivity tensor components in real frequency due to the bulk electronic structure (Eqs. (3-5) in the main text) as calculated using the Kubo formula (Eq.  6 in the main text). As emphasized in the main text, a cutoff energy $\epsilon_{c}$, which describes the validity of the linear energy dispersion of the Weyl Hamiltonian, is introduced in the calculations. We also take that optical transitions do not occur with frequencies greater than $2\epsilon_{c}$, which leads to simplifications especially in the $Im{\sigma_{xy,b}^{inter}}$ \cite{Goswami2013}.

The results from direct evaluation of the optical conductivity components for both types of WSMs are given below.

{\bf Type I WSM; Bulk Interband Components:}
\begin{eqnarray}
\Real{\sigma_{xx,b}^{inter,I}(\Omega)} & = & \frac{\alpha c}{16\pi \hbar v_{F}} \left\lbrace 
\left[\frac{ - 2\tilde{\mu}^{3}}{3\Psi^{3}\Omega^{2}} + \frac{\tilde{\mu}^{2}}{\Psi^{3}\Omega} - \frac{\tilde{\mu}\left( \Psi^{2} + 1\right)}{2\Psi^{3}} + \frac{\left( 4\Psi^{3} + 3\Psi^{2} + 1 \right)\Omega}{12\Psi^{3}}\right]
\chi_{\left[ \Omega_{U}, \Omega_{L} \right]}(\Omega)
 + \frac{2}{3}\Omega
\chi_{\left[ \Omega_{L}, 2\epsilon_{c} \right]}(\Omega) \right\rbrace\nonumber\\
\end{eqnarray}
\begin{eqnarray}
\Real {\sigma_{zz,b}^{inter,I}(\Omega)} & = & \frac{\alpha c}{16\pi \hbar v_{F}} \left\lbrace
\left[\frac{4\tilde{\mu}^{3}}{3\Psi^{3}\Omega^{2}} - \frac{2\tilde{\mu}^{2}}{\Psi^{3}\Omega} - \frac{\tilde{\mu}\left( \Psi^{2} - 1 \right)}{\Psi^{3}}+\frac{\left( 2\Psi^{3} + 3\Psi^{2} - 1\right)\Omega}{6\Psi^{3}}\right]
\chi_{\left[ \Omega_{U}, \Omega_{L} \right]}(\Omega)
 + \frac{2}{3}\Omega
\chi_{\left[ \Omega_{L}, 2\epsilon_{c} \right]}(\Omega) \right\rbrace\nonumber\\
\end{eqnarray}
\begin{eqnarray}
\Imag{\sigma_{xy,b}^{inter}(\Omega)} = \frac{\alpha c}{16\pi\hbar v_{F}}
\Bigg[ & &
\frac{\eta}{\Psi^{2}}\left( \tilde{\mu} - \frac{\tilde{\mu}^{2}}{\Omega} - \frac{\left( 1 - \Psi^{2}\right)\Omega}{4} \right)
\chi_{\left[ \Omega_{U}, \Omega_{L}\right]}(\Omega)
 + \frac{4\epsilon_{c} Q_{z}}{\Omega}
\chi_{\left[ 2\epsilon_{c}, \infty\right]}(\Omega)
\Bigg]
\end{eqnarray}

{\bf Type I WSM; Bulk Intraband Components:}
\begin{eqnarray}
\sigma_{xx,b}^{intra,I}(\Omega) & = & \frac{\ii}{\Omega+\ii\hbar\Gamma}\frac{\alpha c}{4\pi^{2} \Psi^{3}} \frac{\tilde{\mu}^2}{\hbar v_{F}} \left(\frac{\Psi }{1-\Psi^{2}}-\tanh ^{-1}(\Psi )\right)\\
\sigma_{zz,b}^{intra,I}(\Omega) & = & \frac{\ii}{\Omega+\ii\hbar\Gamma}\frac{\alpha c}{2\pi^{2}}\frac{\tilde{\mu}^2}{\hbar v_{F}}\frac{\tanh ^{-1}(\Psi ) - \Psi}{\Psi^{3}}
\end{eqnarray}

{\bf Type II WSM; Bulk Interband Components:}
\begin{eqnarray}
\Real {\sigma_{xx,b}^{inter,II}(\Omega)} = \frac{\alpha c}{16\pi \hbar v_{F}}\Bigg\lbrace & &
\left[\frac{ - 2\tilde{\mu}^{3}}{3\Psi^{3}\Omega^{2}} + \frac{\tilde{\mu}^{2}}{\Psi^{3}\Omega} - \frac{\tilde{\mu}\left( \Psi^{2} + 1\right)}{2\Psi^{3}} + \frac{\left( 4\Psi^{3} + 3\Psi^{2} + 1 \right) \Omega}{12\Psi^{3}}\right]
\chi_{\left[ \Omega_{U}, \Omega_{L} \right]}(\Omega)\nonumber\\
& & +
\left[\frac{2\tilde{\mu}^{2}}{\Psi^{3}\Omega} + \frac{\left( 3\Psi^{2} + 1\right)\Omega}{6\Psi^{3}}\right]
\chi_{\left[ \Omega_{L}, 2\epsilon_{c} \right]}(\Omega) \Bigg\rbrace
\end{eqnarray}
\begin{eqnarray}
\Real {\sigma_{zz,b}^{inter,II}(\Omega)} = \frac{\alpha c}{16\pi \hbar v_{F}}\Bigg\lbrace & &
\left[ \frac{4\tilde{\mu}^{3}}{3\Psi^{3}\Omega^{2}} - \frac{2\tilde{\mu}^{2}}{\Psi^{3}\Omega} - \frac{\tilde{\mu}\left( \Psi^{2} - 1 \right)}{\Psi^{3}} + \frac{\left( 2\Psi^{3} + 3\Psi^{2} - 1\right)\Omega}{6\Psi^{3}}\right]
\chi_{\left[ \Omega_{U}, \Omega_{L} \right]}(\Omega)\nonumber\\
& & + \left[\frac{ - 4\tilde{\mu}^{2}}{\Psi^{3}\Omega} + \frac{\left( 3\Psi^{2} - 1 \right)\Omega}{3\Psi^{3}}\right]
\chi_{\left[ \Omega_{L}, 2\epsilon_{c} \right]}(\Omega) \Bigg\rbrace
\end{eqnarray}
\begin{eqnarray}
\Imag{\sigma_{xy,b}^{inter}(\Omega)} = \frac{\alpha c}{16\pi\hbar v_{F}}
\Bigg\lbrace & &
\frac{\eta}{\Psi^{2}}\left( \tilde{\mu} - \frac{\tilde{\mu}^{2}}{\Omega} + \frac{\left(  \Psi^{2} - 1\right)\Omega}{4} \right)
\chi_{\left[ \Omega_{U}, \Omega_{L}\right]}(\Omega) + 
\left(\frac{2\eta\tilde{\mu}}{\Psi^{2}}\right)
\chi_{\left[ \Omega_{L}, 2\epsilon_{c}\right]}(\Omega)
\nonumber\\
& & + 
\left(\frac{4\epsilon_{c}Q_{z}}{\Omega}\right)
\chi_{\left[ 2\epsilon_{c}, \infty \right]}(\Omega)
\Bigg\rbrace
\end{eqnarray}

{\bf Type II WSM; Bulk Intraband Contributions:}
\begin{eqnarray}
\sigma_{xx,b}^{intra,II}(\Omega)  & = &  \frac{\ii}{\Omega+\ii\hbar\Gamma}\frac{\alpha c}{8\pi^{2}}\frac{1}{\Psi^{3} \hbar v_{F}}\Bigg[
\tilde{\mu}^2 \left(\frac{\Psi ^4+\Psi^{2}}{\Psi^{2}-1}-\left(\log \left[\left(\frac{\Psi  \epsilon_{c}}{\tilde{\mu}}\right)^2-\left(1-\frac{\Psi  Q_{z}}{\tilde{\mu}}\right)^2\right] +\log \left[\Psi^{2}-1\right]\right)\right)\nonumber\\
& & +\left(\left(\epsilon_{c}^2+Q_{z}^2\right) \Psi^{2} \left(\Psi^{2}-1\right)-2 Q_{z} \tilde{\mu} \left(\Psi +\Psi^3\right)\right)
\Bigg]
\end{eqnarray}

\begin{eqnarray}
\sigma_{zz,b}^{intra,II}(\Omega) & = &  \frac{\ii}{\Omega+\ii\hbar\Gamma}\frac{\alpha c}{4\pi^{2}}\frac{1}{\Psi^{2} \hbar v_{F}}\Bigg[
\frac{2\tilde{\mu}^{2}}{\Psi}\left(\log \left[\left(\frac{\Psi  \epsilon_{c}}{\tilde{\mu}}\right)^2-\left(1-\frac{\Psi  Q_{z}}{\tilde{\mu}}\right)^2\right] + \log \left[\Psi^{2}-1\right]\right)\nonumber\\
& & + 2 Q_{z} \left(\tilde{\mu}+2 \epsilon_{c} \Psi^{3}-\tilde{\mu} \Psi ^4\right)+Q_{z}^2 \left(\Psi +\Psi ^5\right)+\frac{\Psi  \left(\epsilon_{c}^2 (\Psi -1)^2 \left(1+\Psi ^4\right)+\tilde{\mu}^2
   \left(1-2 \Psi -2 \Psi^{3}+\Psi ^4\right)\right)}{(\Psi -1)^2}
\Bigg]
\end{eqnarray}

In the above expressions, we take that $\alpha = \frac{e^{2}}{\hbar c} \approx \frac{1}{137}$ is the fine structure constant, $\Theta(x)$ is the Heaviside Theta function, and $\chi_{\left[ a, b \right]}(x)$ is the indicator function with $\chi_{\left[ a, b \right]}(x) = 1$ if $x\in\left[ a, b \right]$ and zero otherwise. We have also made the following notation \cite{Mukherjee-18}
\begin{equation}
\Omega = \hbar\omega;  \tilde{\mu} = |\mu - b_{0}|; \Omega_{U}  =  \frac{2\tilde{\mu}}{|\Psi + 1|}; \Omega_{L}  =  \frac{2\tilde{\mu}}{|\Psi - 1 |}; 
\epsilon_{c}  =  \hbar v_{F} k_{c}; Q_{z}  =  \eta\hbar v_{F} b
\end{equation}

The above results show that the bulk conductivity components depend explictly on the cutoff energy $\epsilon_c$, as shown below. Typically one takes the limit $\epsilon_{c}\to\infty$ as done in 2D Dirac materials, for example \cite{Rodriguez-Lopez-18}. Such a limit leads to undesirable consequences in 3D Weyl materials, including singularities in the  Kramers-Krönig relations and disappearance of the topological Hall conductivity. The bulk Hall conductivity has a contribution associated with the nontrivial topology of the WSM and a contribution that is related to other characteristics of the band structure, such as tilting. The topological part is calculated from the Berry curvature, thus this it is independent of $\mu$, $b_{0}$ and $\Psi$, as we found directly from the Kubo formula. 

A graphical representation for the various components of the bulk optical conductivity as a function of frequency for both types of WSMs is given in Fig. 2 in the main text.


\section{Bulk Optical Conductivity Tensor Components at Imaginary frequencies}
Here we summarize our results for the nonzero bulk conductivity tensor components at imaginary frequencies $\omega=\ii\xi$. These are obtained from Kramers-Kr\"onig relations, which are given in Eq. (12) in the main text. We note that taking the limit $\epsilon_{c}\to\infty$ results in divergencies in the  Kramers-Kr\"onig relations. This comes from the fact that $\sigma_{ij}(\omega)$ does not decay at least as $1/\omega$ for large $\omega$ as required for the fullfilment of the Kramers-Kr\"onig relations. Thus the cutoff energy plays a crucial role for the optical response properties of these materials described by the  effective model for the 3D Dirac spectra in Eqs. (1-5) in the main text.

After the direct substitution $\omega\to\ii\xi$ in Eqs. (1-10), we find

{\bf Type I WSM; Bulk Interband Components:}
\begin{eqnarray}
\sigma_{xx,b}^{inter,I}(\tilde{\xi}) = \frac{1}{192 \pi^{2} \Psi^{3} \tilde{\xi} \hbar v_{F}}
& \Bigg[ &
4\frac{\tilde{\mu}}{\tilde{\xi}} \left(4 \tilde{\mu}^2-3 \left(\Psi^{2}+1\right) \tilde{\xi}^{2}\right) \left(\tan ^{-1}\left(\frac{2 \tilde{\mu}}{(1-\Psi ) \tilde{\xi} }\right)-\tan ^{-1}\left(\frac{2 \tilde{\mu}}{(\Psi +1) \tilde{\xi} }\right)\right)
\nonumber\\ & &
+ \left(\left((3-4 \Psi ) \Psi^{2}+1\right) \tilde{\xi}^{2}-12 \tilde{\mu}^2\right) \log \left(\frac{4 \tilde{\mu}^2}{(\Psi -1)^2}+\tilde{\xi}^{2}\right)
\nonumber\\ & &
-\left( \left((3 + 4\Psi) \Psi^{2}+1\right) \tilde{\xi}^{2} - 12 \tilde{\mu}^2\right) \log \left(\frac{4 \tilde{\mu}^2}{(\Psi +1)^2}+\tilde{\xi}^{2}\right)
\nonumber\\ & &
-8 \tilde{\mu}^2 \left(2 \Psi +3 \log \left(\frac{1-\Psi }{1 + \Psi}\right)\right)+8 \Psi^{3} \tilde{\xi}^{2} \log \left(4 \epsilon_{c}^2+\tilde{\xi}^{2}\right)
\Bigg],
\end{eqnarray}
\begin{eqnarray}
\sigma_{zz,b}^{inter,I}(\tilde{\xi}) = \frac{1}{96 \pi^{2} \Psi^{3} \tilde{\xi} \hbar v_{F}}
& \Bigg[ &
4 \frac{\tilde{\mu}}{\tilde{\xi} } \left(4 \tilde{\mu}^2+3 \left(\Psi^{2}-1\right) \tilde{\xi}^{2}\right) \left(\tan ^{-1}\left(\frac{2 \tilde{\mu}}{(\Psi -1) \tilde{\xi} }\right)+\tan ^{-1}\left(\frac{2 \tilde{\mu}}{(\Psi +1) \tilde{\xi} }\right)\right)
\nonumber\\ & &
+ \left(12 \tilde{\mu}^2+(2 \Psi -1) (\Psi +1)^2 \tilde{\xi}^{2}\right) \left(\log \left(\frac{4 \tilde{\mu}^2}{(\Psi -1)^2}+\tilde{\xi}^{2}\right)-\log \left(\frac{4 \tilde{\mu}^2}{(\Psi +1)^2}+\tilde{\xi}^{2}\right)\right)
\nonumber\\ & &
+16 \tilde{\mu}^2 \Psi -48 \tilde{\mu}^2 \tanh ^{-1}(\Psi )+4 \Psi^{3} \tilde{\xi}^{2} \left(\log \left(4 \epsilon_{c}^2+\tilde{\xi}^{2}\right)-\log \left(\frac{4 \tilde{\mu}^2}{(\Psi -1)^2}+\tilde{\xi}^{2}\right)\right)
\Bigg],
\end{eqnarray}
\begin{eqnarray}
\sigma_{xy,b}^{inter}(\tilde{\xi}) = \frac{\eta }{32 \pi^{2} \tilde{\xi}  \hbar v_{F}}
& \Bigg[ &
16 Q_{z} \epsilon_{c} \left(\frac{\pi }{2}-\tan ^{-1}\left(\frac{2 \epsilon_{c}}{\tilde{\xi} }\right)\right)
\nonumber\\ & &
+\frac{1}{\Psi^{2}}\left(4 \tilde{\mu}^2+\left(\Psi^{2}-1\right) \tilde{\xi}^{2}\right) \left(\tan ^{-1}\left(\frac{2 \tilde{\mu}}{(\Psi -1) \tilde{\xi} }\right)+\tan ^{-1}\left(\frac{2 \tilde{\mu}}{(\Psi +1) \tilde{\xi} }\right)\right)
\nonumber\\ & &
-\frac{2\tilde{\mu}\tilde{\xi}}{\Psi^{2}}\left( 2\Psi -\log \left(\frac{4 \tilde{\mu}^2}{(\Psi -1)^2}+\tilde{\xi}^{2}\right)+\log \left(\frac{4 \tilde{\mu}^2}{(\Psi +1)^2}+\tilde{\xi}^{2}\right) \right)
\Bigg].
\end{eqnarray}

{\bf Type II WSM; Bulk Interband Components:}
\begin{eqnarray}
\sigma_{xx,b}^{inter,II}(\tilde{\xi}) = \frac{1}{192 \pi^{2} \Psi^{3} \tilde{\xi} \hbar v_{F}}
& \Bigg[ &
4 \frac{\tilde{\mu}}{\tilde{\xi}} \left(4 \tilde{\mu}^2-3 \left(\Psi^{2}+1\right) \tilde{\xi}^{2}\right) \left(\tan ^{-1}\left(\frac{2 \tilde{\mu}}{(\Psi -1) \tilde{\xi} }\right)-\tan ^{-1}\left(\frac{2 \tilde{\mu}}{(\Psi +1) \tilde{\xi} }\right)\right)
\nonumber\\ & &
+ \left(\left((4 \Psi +3) \Psi^{2}+1\right) \tilde{\xi}^{2}-12 \tilde{\mu}^2\right) \left(\log \left(\frac{4 \tilde{\mu}^2}{(\Psi -1)^2}+\tilde{\xi}^{2}\right)-\log \left(\frac{4 \tilde{\mu}^2}{(\Psi +1)^2}+\tilde{\xi}^{2}\right)\right)
\nonumber\\ & &
 - 2 \left( \left(3 \Psi^{2}+1\right) \tilde{\xi}^{2} - 12 \tilde{\mu}^2\right) \left( \log \left(\frac{4 \tilde{\mu}^2}{(\Psi -1)^2}+\tilde{\xi}^{2}\right) - \log \left(4 \epsilon_{c}^2+\tilde{\xi}^{2}\right)\right)
\nonumber\\ & &
+ 24 \tilde{\mu}^2 \left(2 \log \left(\frac{\epsilon_{c}}{\tilde{\mu}}\right)+\log \left(\Psi^{2}-1\right)\right) -16 \tilde{\mu}^2
\Bigg],
\end{eqnarray}
\begin{eqnarray}
\sigma_{zz,b}^{inter,II}(\tilde{\xi}) = \frac{1}{96 \pi^{2} \Psi^{3} \tilde{\xi} \hbar v_{F}}
& \Bigg[ &
4 \frac{\tilde{\mu}}{\tilde{\xi}} \left(4 \tilde{\mu}^2+3 \left(\Psi^{2}-1\right) \tilde{\xi}^{2}\right) \left(\tan ^{-1}\left(\frac{2 \tilde{\mu}}{(\Psi +1) \tilde{\xi} }\right)-\tan ^{-1}\left(\frac{2 \tilde{\mu}}{(\Psi -1) \tilde{\xi} }\right)\right)
\nonumber\\ & &
+ \left( (2 \Psi -1) (\Psi +1)^2 \tilde{\xi}^{2} + 12 \tilde{\mu}^2 \right) \left(\log \left(\frac{4 \tilde{\mu}^2}{(\Psi -1)^2}+\tilde{\xi}^{2}\right)-\log \left(\frac{4 \tilde{\mu}^2}{(\Psi +1)^2}+\tilde{\xi}^{2}\right)\right)
\nonumber\\ & &
+ 2 \left( \left(3 \Psi^{2}-1\right) \tilde{\xi}^{2} + 12 \tilde{\mu}^2\right) \left(\log \left(4 \epsilon_{c}^2+\tilde{\xi}^{2}\right)-\log \left(\frac{4 \tilde{\mu}^2}{(\Psi -1)^2}+\tilde{\xi}^{2}\right)\right)
\nonumber\\ & &
+ 8 \tilde{\mu}^2 \left( 2 -6 \log \left(\frac{\epsilon_{c}}{\tilde{\mu}}\right)-3 \log \left(\Psi^{2}-1\right)\right)
\Bigg],
\end{eqnarray}
\begin{eqnarray}
\sigma_{xy,b}^{inter}(\tilde{\xi}) = \frac{\eta }{8 \pi^{2} \Psi^{2} \tilde{\xi}  \hbar v_{F}}
& \Bigg[ &
4 Q_{z} \Psi^{2} \epsilon_{c} \left(\frac{\pi }{2}-\tan ^{-1}\left(\frac{2 \epsilon_{c}}{\tilde{\xi} }\right)\right) + \tilde{\mu}\tilde{\xi}
\nonumber\\ & &
+ \left(\tilde{\mu}^2+\frac{1}{4} \left(\Psi^{2}-1\right) \tilde{\xi}^{2}\right) \left(\tan ^{-1}\left(\frac{2 \tilde{\mu}}{(\Psi +1) \tilde{\xi} }\right)-\tan ^{-1}\left(\frac{2 \tilde{\mu}}{(\Psi -1) \tilde{\xi} }\right)\right)
\nonumber\\ & &
- \frac{\tilde{\mu}\tilde{\xi}}{2}\left(\log \left(\frac{4 \tilde{\mu}^2}{(\Psi -1)^2}+\tilde{\xi}^{2}\right)+\log \left(\frac{4 \tilde{\mu}^2}{(\Psi +1)^2}+\tilde{\xi}^{2}\right)-2 \log \left(4 \epsilon_{c}^2+\tilde{\xi}^{2}\right)\right)
\Bigg].
\end{eqnarray}
where we have defined $\tilde{\xi} = \hbar(\xi + \Gamma)$.

A graphical representation for the various components of the bulk optical conductivity as a function of imaginary frequency for both types of WSMs is given in Fig. 3 in the main text.

\newpage
\section{Fresnel coefficients for a 3D semi-infinite space occupied by a material with anisotropic optical response properties and 2D surface currents}
\begin{figure}[H]
\centering
\includegraphics[width=0.45\linewidth]{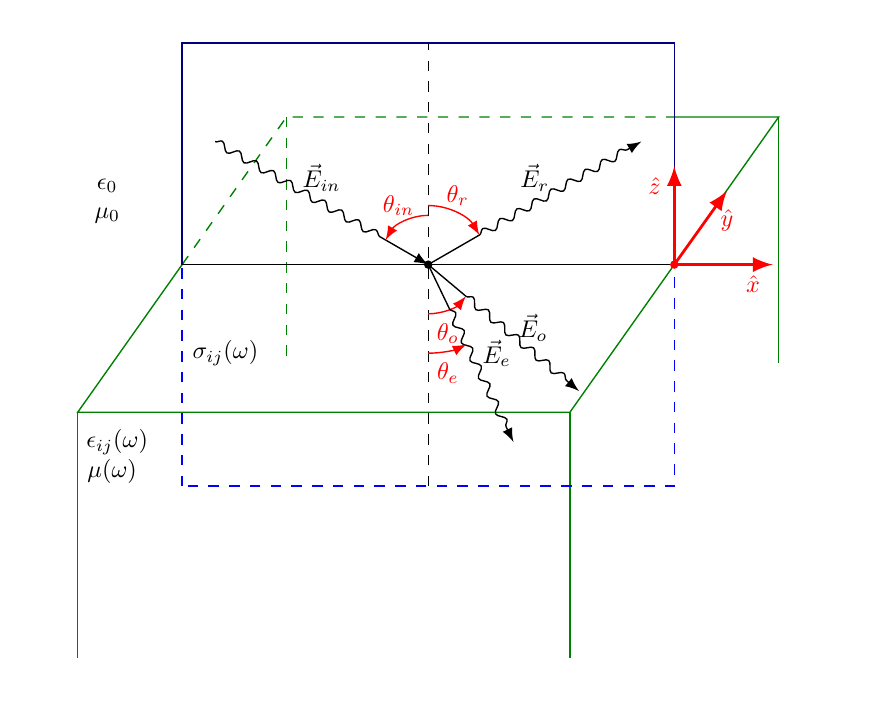}
\caption{ (Color online) Diagram of the reflection and transmitted electric fields from a planar media with anistropic response and surface conductivities. We observe that the field at the vacuum is the sum of the incident and reflected fields, and that the transmitted field is the (possible) sum of two different propagating modes, the ordinary and extraordinary.}
\label{fig:Diagram_Reflection_anisotropic_media}
\end{figure}

Solving the Maxwell's equations together with the electromagnetic boundary conditions results in the Fresnel reflection coefficients, which are necessary for the Casimir interaction calculations. Here we work in Cartesian coordinates for anisotropic medium taking up the $z<0$ space. The boundary condition for the electric $\bm{E}_{\text{vac, diel}}$ and magnetic $\bm{H}_{\text{vac, die}}$ fields in the vacuum (vac) $z>0$ and dielectric (die) $z<0$ regions  are given in the presence of surface currents described by Ohm's law 
\begin{eqnarray}\label{EMBoundaryCondition}
\hat{\bm{z}}\times[\bm{E}_{\text{vac}} - \bm{E}_{\text{die}}] & = & 0\nonumber\\
\label{Magnetic_BC}\hat{\bm{z}}\times[\bm{H}_{\text{vac}} - \bm{H}_{\text{die}}] & = & \frac{4\pi}{c}\bm{J}_{s} = \frac{4\pi}{c}\sigma_{s}\bm{E}_{\text{die}},
\end{eqnarray}
where $\hat{\bm{z}}$ is the unit vector perpendicular to the $z=0$ plane. We can observe a diagram of the system under study in \Fig{fig:Diagram_Reflection_anisotropic_media}.

By taking advantage of the planar geometry, the solution to the Maxwell's equations is sought in terms of the following orthogonal functions \cite{Rahi2009}
 \begin{eqnarray}
\bm{M}_{\bm{k}}(\bm{r}) & = & \frac{1}{k_{\parallel}}\nabla\times\left[\hat{\bm{z}}\phi_{\bm{k}}(\bm{r})\right]\nonumber\\
\bm{N}_{\bm{k}}(\bm{r}) & = & \frac{c}{\omega}\nabla\times\bm{M}_{\bm{k}}(\bm{r}),
\end{eqnarray}
where $\phi_{\bm{k}}(\bm{r}) = e^{\ii\left(\bm{k}\cdot\bm{r} - \omega t\right)}$, $\bm{k} = \left(k_{x},k_{y},k_{z}\right)$ is the wave vector, and $k_{\parallel} = \sqrt{ k_{x}^{2} + k_{y}^{2} }$.  The property $\bm{M}_{\bm{k}}(\bm{r}) = \frac{c}{\omega}\nabla\times\bm{N}_{\bm{k}}(\bm{r})$ makes this representation convenient in realizing that the electromagnetic wave propagation can be decoupled as $s$ and $p$ polarized fields.

{\it Incident Fields} The general incident electric field will be the superposition of those two polarized waves
\begin{eqnarray}\label{DefEincident}
\bm{E}_{\text{in}} & = & A_{\perp}\bm{M}_{\bm{k}}^{\text{in}}(\bm{r}) + A_{\parallel}\bm{N}_{\bm{k}}^{\text{in}}(\bm{r}).
\end{eqnarray}
From the Maxwell Equations, the magnetic field can be obtained by pluging the result of \Eq{DefEincident} in $\nabla\times\bm{E} = \frac{-1}{c}\frac{\partial\bm{B}}{\partial t}$ with the constitutive relation of the magnetic field in vacuum $\bm{B} = \bm{\mu}^{-1}\cdot\bm{H} = \mu_{0}^{-1}\bm{H}$, then
\begin{eqnarray}\label{DefHincident}
\bm{H}_{\text{in}} & = & -\ii \left[ A_{\parallel}\bm{M}_{\bm{k}}^{\text{in}}(\bm{r}) + A_{\perp}\bm{N}_{\bm{k}}^{\text{in}}(\bm{r})\right].
\end{eqnarray}
where $A_{\perp}$ and $A_{\parallel}$ are coefficients and 
\begin{eqnarray}
\bm{M}_{\bm{k}}^{\text{in}}(\bm{r}) = \frac{1}{k_{\parallel}}\ii\, e^{\ii\left(\bm{k}\cdot\bm{r} - \omega t\right)}\left(\begin{array}{c}
\phantom{+}k_{y}\\
- k_{x}\\
0
\end{array}
\right),
\hspace{2cm}
\bm{N}_{\bm{k}}^{\text{in}}(\bm{r}) = \frac{1}{k_{\parallel}}\frac{c}{\omega} e^{\ii\left(\bm{k}\cdot\bm{r} - \omega t\right)}\left(\begin{array}{c}
- k_{x}k_{z}\\
- k_{y}k_{z}\\
k_{x}^{2} + k_{y}^{2}
\end{array}
\right).
\end{eqnarray}

{\it Reflected Fields} From Snell's law it follows that for the wave vector of incidence $\bm{k} = \left(k_{x},k_{y},k_{z}\right)$ at the $z=0$ plane, the reflected wave vector is $\bm{k}_{r} = \left(k_{x},k_{y}, - k_{z}\right)$. Thus the reflected electric and magnetic fields can be written in a similar way, 
\begin{eqnarray}\label{DefEHreflected}
\bm{E}_{\text{r}} & = & \phantom{+1}R_{\perp}\bm{M}_{\bm{k}}^{\text{r}}(\bm{r}) + R_{\parallel}\bm{N}_{\bm{k}}^{\text{r}}(\bm{r}),\nonumber\\
\bm{H}_{\text{r}} & = & -\ii \left[ R_{\parallel}\bm{M}_{\bm{k}}^{\text{r}}(\bm{r}) + R_{\perp}\bm{N}_{\bm{k}}^{\text{r}}(\bm{r})\right].
\end{eqnarray}
where the orthogonal functions are
\begin{eqnarray}
\bm{M}_{\bm{k}}^{\text{r}}(\bm{r}) = \frac{1}{k_{\parallel}}\ii\, e^{\ii\left(\bm{k}_{r}\cdot\bm{r} - \omega t\right)}\left(\begin{array}{c}
\phantom{+}k_{y}\\
- k_{x}\\
0
\end{array}
\right),
\hspace{2cm}
\bm{N}_{\bm{k}}^{\text{r}}(\bm{r}) = \frac{1}{k_{\parallel}}\frac{c}{\omega} e^{\ii\left(\bm{k}_{r}\cdot\bm{r} - \omega t\right)}\left(\begin{array}{c}
k_{x}k_{z}\\
k_{y}k_{z}\\
k_{x}^{2} + k_{y}^{2}
\end{array}
\right).
\end{eqnarray}

{\it Transmitted Fields} To obtain $\bm{E}_{\text{r}}$ and $\bm{H}_{\text{r}}$, the Maxwell equations are solved for planar wave with $\bm{E} = \bm{\mathcal{E}}e^{\ii(\bm{k}\cdot\bm{r} - \omega t)},\,\,\bm{H} = \bm{\mathcal{H}}e^{\ii(\bm{k}\cdot\bm{r} - \omega t)}$ in a medium with $\mu=\mu_0 \hat{1}$ with anisotropic dielectric function given as
 \begin{eqnarray}
\bm{\epsilon} & = & \left(\begin{array}{ccc}
\epsilon_{xx} & \epsilon_{xy} & \epsilon_{xz}\\
\epsilon_{yx} & \epsilon_{yy} & \epsilon_{yz}\\
\epsilon_{zx} & \epsilon_{zy} & \epsilon_{zz}
\end{array}
\right).
\end{eqnarray}
Taking into account the constitutive relations $\bm{D} = \bm{\epsilon}\cdot\bm{E}$ and $\bm{H} = \bm{\mu}^{-1}\cdot\bm{B}$ and using $\nabla\times\bm{E} = \frac{-1}{c}\bm{\mu}\cdot\partial_{t}\bm{H}$ and $\nabla\times\bm{H} = \frac{+1}{c}\bm{\epsilon}\cdot\partial_{t}\bm{E}$, one finds in particular

\begin{eqnarray}\label{SolutionHz}
\mathcal{H}_{z} & = & \frac{c}{\mu\omega}\left( k_{x}\mathcal{E}_{y} - k_{y}\mathcal{E}_{x} \right),\\
\mathcal{E}_{z} & = & \frac{-1}{\epsilon_{zz}}\left[
\frac{c}{\omega}\left( k_{x}\mathcal{H}_{y} - k_{y}\mathcal{H}_{x} \right) + 
\epsilon_{xx}\mathcal{E}_{x} + \epsilon_{xy}\mathcal{E}_{y}.
\right]\label{SolutionEz}
\end{eqnarray}
We further note that the rest of the $\bm{\mathcal{E}}$ and $\bm{\mathcal{H}}$ components can be found by organizing the equations $\nabla\times\bm{E} = \frac{-1}{c}\bm{\mu}\cdot\partial_{t}\bm{H}$ and $\nabla\times\bm{H} = \frac{+1}{c}\bm{\epsilon}\cdot\partial_{t}\bm{E}$ in the following matrix form equation
\begin{eqnarray}
\bm{A}\cdot\bm{v} & = & \bm{0},
\end{eqnarray}
where 
\begin{equation}
\bm{v}^{\top} = (\mathcal{E}_{x},\mathcal{E}_{y},\mathcal{H}_{x},\mathcal{H}_{y}), 
\end{equation}
\begin{eqnarray}
\bm{A} = \left(
\begin{array}{cccc}
 \frac{c^{2}\left[ k_{y}^{2}\epsilon^{-1}_{xx} - k_{x} k_{y}\epsilon^{-1}_{xy} \right]}{\mu\omega^{2}} - 1 & \frac{c^{2} \left[ k_{x}^{2}\epsilon^{-1}_{xy} - k_{x} k_{y}\epsilon^{-1}_{xx}\right]}{\mu\omega^{2}} & \frac{c\left[ k_{y} \epsilon^{-1}_{xz} - k_{z} \epsilon^{-1}_{xy} \right]}{\omega} & \frac{c\left[ k_{z} \epsilon^{-1}_{xx} - k_{x} \epsilon^{-1}_{xz} \right]}{\omega}\\
 \frac{c^{2}\left[ k_{y}^{2}\epsilon^{-1}_{yx} - k_{x} k_{y} \epsilon^{-1}_{yy} \right]}{\mu\omega^{2}} & \frac{c^{2}\left[ k_{x}^2 \epsilon^{-1}_{yy} - k_{x} k_{y}\epsilon^{-1}_{yx} \right]}{\mu\omega^{2}} - 1 & \frac{c\left[ k_{y}\epsilon^{-1}_{yz} - k_{z}\epsilon^{-1}_{yy} \right]}{\omega} & \frac{c\left[ k_{z}\epsilon^{-1}_{yx} - k_{x} \epsilon^{-1}_{yz} \right]}{\omega}\\
\frac{c^{3}\left[ k_{y}^{3}\epsilon^{-1}_{zx} - k_{x} k_{y}^{2}\epsilon^{-1}_{zy} \right]}{\mu^{2}\omega^{3}} & \frac{c^{3}\left[ k_{x}^{2}k_{y}\epsilon^{-1}_{zy} - k_{x}k_{y}^{2}\epsilon^{-1}_{zx} \right]}{\mu^{2}\omega^{3}} - \frac{ck_{z}}{\mu\omega} & \frac{c^{2}\left[ k_{y}^{2}\epsilon^{-1}_{zz} - k_{y} k_{z}\epsilon^{-1}_{zy} \right]}{\mu\omega^{2}} - 1 & \frac{c^{2}\left[ k_{y}k_{z}\epsilon^{-1}_{zx} - k_{x}k_{y}\epsilon^{-1}_{zz} \right]}{\mu\omega^{2}}\\
\frac{c^{3}\left[ k_{x}^{2}k_{y}\epsilon^{-1}_{zy}  - k_{x}k_{y}^{2}\epsilon^{-1}_{zx} \right]}{\mu^{2}\omega^{3}} + \frac{ck_{z}}{\mu\omega} & \frac{c^{3}\left[ k_{x}^{2}k_{y}\epsilon^{-1}_{zx} - k_{x}^{3}\epsilon^{-1}_{zy} \right]}{\mu^{2}\omega^{3}} &  \frac{c^{2}\left[ k_{x}k_{z}\epsilon^{-1}_{zy} - k_{x}k_{y}\epsilon^{-1}_{zz} \right]}{\mu\omega^{2}} & \frac{c^{2}\left[ k_{x}^{2}\epsilon^{-1}_{zz} - k_{x}k_{z}\epsilon^{-1}_{zx} \right]}{\mu\omega^{2}} - 1\\
\end{array}
\right).
\end{eqnarray}
where $\epsilon^{-1}_{ij}$ is the $ij-$component of the inverse of $\bm{\epsilon}$. Eqs. (28-30) can further be re-written as an eigenvalue problem, such that 
\begin{eqnarray}
\bm{M}\cdot\bm{v}_{n} & = & k_{z,n}\bm{v}_{n}
\end{eqnarray}
with
\begin{eqnarray}
\bm{M} = \left(
\begin{array}{cccc}
 - k_{x}\frac{\epsilon_{zx}}{\epsilon_{zz}} & -k_{x}\frac{\epsilon_{zy}}{\epsilon_{zz}} & \frac{c k_{x} k_{y}}{\omega\epsilon_{zz}} & \frac{\mu\omega}{c} - \frac{c k_{x}^2}{\omega\epsilon_{zz}} \\
 - k_{y}\frac{\epsilon_{zx}}{\epsilon_{zz}} & - k_{y}\frac{\epsilon_{zy}}{\epsilon_{zz}} & - \frac{\mu\omega}{c} + \frac{c k_{y}^{2}}{\omega\epsilon_{zz}} & -\frac{c k_{x} k_{y}}{\omega\epsilon_{zz}} \\
- \frac{c k_{x} k_{y}}{\mu\omega} + \frac{\omega}{c\epsilon_{zz}}\left(\epsilon_{yz} \epsilon_{zx} - \epsilon_{yx} \epsilon_{zz}\right)  & \frac{c k_{x}^{2}}{\mu\omega} + \frac{\omega}{c\epsilon_{zz}}\left(\epsilon_{yz} \epsilon_{zy}-\epsilon_{yy} \epsilon_{zz}\right) & -k_{y}\frac{\epsilon_{yz}}{\epsilon_{zz}} & k_{x}\frac{\epsilon_{yz}}{\epsilon_{zz}} \\
- \frac{c k_{y}^{2}}{\mu\omega} + \frac{\omega}{c\epsilon_{zz}}\left(\epsilon_{xx} \epsilon_{zz} - \epsilon_{xz}\epsilon_{zx}\right)  & \frac{c k_{x} k_{y}}{\mu\omega} + \frac{\omega}{c\epsilon_{zz}}\left(\epsilon_{xy} \epsilon_{zz}-\epsilon_{xz} \epsilon_{zy}\right) & k_{y}\frac{\epsilon_{xz}}{\epsilon_{zz}} & -k_{x}\frac{\epsilon_{xz}}{\epsilon_{zz}} \\
\end{array}
\right).
\end{eqnarray}
The eigenfunctions of $\bm{M}$ with $k_{z}<0$ eigenvalue correspond to waves propagating to $z\to-\infty$ and are the allowed solutions for the transmitted waves. As a result, there are two different solutions (degenerated when $\bm{\epsilon}$ is isotropic): $\bm{v}_{o}$ with $k_{z,o} = q_{o}$ and $\bm{v}_{e}$ with $k_{z,e} = q_{e}$. The corresponding fields can then be written as
\begin{eqnarray}
\bm{E}_{o}(\bm{k}) = \left(\begin{array}{c}
v_{o,1}\\
v_{o,2}\\
E_{o,z}
\end{array}\right)e^{ \ii\left( \bm{k}_{o}\cdot\bm{r} - \omega t\right) },
\hspace{2cm}
\bm{H}_{o}(\bm{k}) = \left(\begin{array}{c}
v_{o,3}\\
v_{o,4}\\
H_{o,z}
\end{array}\right)e^{ \ii\left( \bm{k}_{o}\cdot\bm{r} - \omega t\right) },
\end{eqnarray}
and
\begin{eqnarray}
\bm{E}_{e}(\bm{k}) = \left(\begin{array}{c}
v_{e,1}\\
v_{e,2}\\
E_{e,z}
\end{array}\right)e^{ \ii\left( \bm{k}_{e}\cdot\bm{r} - \omega t\right) },
\hspace{2cm}
\bm{H}_{e}(\bm{k}) = \left(\begin{array}{c}
v_{e,3}\\
v_{e,4}\\
H_{e,z}
\end{array}\right)e^{ \ii\left( \bm{k}_{e}\cdot\bm{r} - \omega t\right) },
\end{eqnarray}
where $\bm{k}_{\text{o}} = (k_{x},k_{y},q_{o})$, $\bm{k}_{\text{e}} = (k_{x},k_{y},q_{e})$, and the $z-$components of $(\mathcal{E}_{z}, \mathcal{H}_{z})$ can be obtained from the $x$ and $y$ components of $\mathcal{E}_{z}$ and $\mathcal{H}_{z}$ that appear in $\bm{v}$.

\subsection{General Fresnel coefficients}
The Fresnel reflection $\mathbb{R}_{\bm{k},\bm{q}}$ and transmission $\mathbb{T}_{\bm{k},\bm{q}}$ matrices are defined as operators relating the reflected and transmitted fields to the incident field according to:
\begin{eqnarray}\label{DefinitionRT}
\bm{E}_{\bm{k}}^{\text{r}} & = & \mathbb{R}_{\bm{k},\bm{q}}\bm{E}_{\bm{q}}^{\text{in}}\nonumber\\
\bm{E}_{\bm{k}}^{\text{t}} & = & \mathbb{T}_{\bm{k},\bm{q}}\bm{E}_{\bm{q}}^{\text{in}}.
\end{eqnarray}
Using the multipolar basis in \Eq{DefinitionRT}, we have 
\begin{eqnarray}
\left(\begin{array}{c}
\bm{M}_{\bm{k}}^{\text{r}}\\
\bm{N}_{\bm{k}}^{\text{r}}
\end{array}\right) = \left(\begin{array}{cc}
r_{MM} & r_{MN}\\
r_{NM} & r_{NN}
\end{array}\right)\left(\begin{array}{c}
\bm{M}_{\bm{k}}^{\text{in}}\\
\bm{N}_{\bm{k}}^{\text{in}}
\end{array}\right) & \Leftrightarrow & \left(\begin{array}{c}
R_{\perp}\\
R_{\parallel}
\end{array}\right) = \left(\begin{array}{cc}
r_{MM} & r_{MN}\\
r_{NM} & r_{NN}
\end{array}\right)\left(\begin{array}{c}
A_{\perp}\\
A_{\parallel}
\end{array}\right)\\
\left(\begin{array}{c}
\bm{E}_{\bm{k}}^{\text{o}}\\
\bm{E}_{\bm{k}}^{\text{e}}
\end{array}\right) = \left(\begin{array}{cc}
t_{oM} & t_{oN}\\
t_{eM} & t_{eN}
\end{array}\right)\left(\begin{array}{c}
\bm{M}_{\bm{k}}^{\text{in}}\\
\bm{N}_{\bm{k}}^{\text{in}}
\end{array}\right) & \Leftrightarrow & \left(\begin{array}{c}
T_{\text{o}}\\
T_{\text{e}}
\end{array}\right) = \left(\begin{array}{cc}
t_{oM} & t_{oN}\\
t_{eM} & t_{eN}
\end{array}\right)\left(\begin{array}{c}
A_{\perp}\\
A_{\parallel}
\end{array}\right)
\end{eqnarray}
which specify the components $r_{MM},  r_{MN}, r_{NM}, r_{NN}$ of the reflection matrix and the components $t_{oM}, t_{oN}, t_{eM}, t_{eN}$ of the transmission matrix. The reflection matrices can be directly found from the boundary conditions 
\begin{eqnarray}\label{EMBoundaryConditionPlate}
\hat{\bm{n}}\times[\bm{E}_{\text{vac}} - \bm{E}_{\text{die}}] & = & 0\\
\hat{\bm{n}}\times[\bm{H}_{\text{vac}} - \bm{H}_{\text{die}}] & = & \frac{4\pi}{c}\bm{J}_{s} = \frac{4\pi}{c}\sigma_{s}\bm{E}_{\text{die}}, \nonumber
\end{eqnarray}
where the incident, reflected, and transmitted fields are found from the previous section. The fields in the vacuum and the medium are $\bm{E}_{\text{vac}} = \bm{E}_{\text{in}} + \bm{E}_{\text{r}}$, $\bm{E}_{\text{die}} = T_{o}\bm{E}_{\text{o}} + T_{e}\bm{E}_{\text{e}}$ and a similar expression holds for the magnetic field $\bm{H}$. Note that for isotropic dielectrics, $\bm{E}_{\text{die}} = T_{o}\bm{E}_{\text{o}} + T_{e}\bm{E}_{\text{e}} = T_{o}\bm{M}_{\bm{k}} + T_{e}\bm{N}_{\bm{k}}$, and the full derivation is greatly simplified. 

For the planar geometry here we take $\hat{\bm{n}} = \hat{\bm{z}} = \left(0,0,1\right)$. Applying a scalar product in the above boundary conditions by the two different incident polarized waves ($\bm{M}^{\dagger}_{\bm{k}}$ and $\bm{N}^{\dagger}_{\bm{k}}$), one finds 4 equations with 6 variables, such that 
\begin{eqnarray}
\bm{M}^{\dagger}_{\bm{k}}\cdot\left(\hat{\bm{z}}\times[\bm{E}_{\text{in}} + \bm{E}_{\text{r}} - \left(T_{o}\bm{E}_{\text{o}} + T_{e}\bm{E}_{\text{e}}\right)]\right) & = & 0\nonumber\\
\bm{N}^{\dagger}_{\bm{k}}\cdot\left(\hat{\bm{z}}\times[\bm{E}_{\text{in}} + \bm{E}_{\text{r}} - \left(T_{o}\bm{E}_{\text{o}} + T_{e}\bm{E}_{\text{e}}\right)]\right) & = & 0\nonumber\\
\bm{M}^{\dagger}_{\bm{k}}\cdot\left(\hat{\bm{z}}\times[\bm{H}_{\text{in}} + \bm{H}_{\text{r}} - \left(T_{o}\bm{H}_{\text{o}} + T_{e}\bm{H}_{\text{e}}\right)]\right) & = & \bm{M}^{\dagger}_{\bm{k}}\cdot\left[\frac{4\pi}{c}\sigma_{s}\left(T_{o}\bm{E}_{\text{o}} + T_{e}\bm{E}_{\text{e}}\right)\right]\nonumber\\
\bm{N}^{\dagger}_{\bm{k}}\cdot\left(\hat{\bm{z}}\times[\bm{H}_{\text{in}} + \bm{H}_{\text{r}} - \left(T_{o}\bm{H}_{\text{o}} + T_{e}\bm{H}_{\text{e}}\right)]\right) & = & \bm{N}^{\dagger}_{\bm{k}}\cdot\left[\frac{4\pi}{c}\sigma_{s}\left(T_{o}\bm{E}_{\text{o}} + T_{e}\bm{E}_{\text{e}}\right)\right].
\end{eqnarray}
Substituting the incident and reflected field at the boundary ($z=0$) by \Eq{DefEincident}, \Eq{DefHincident}and \Eq{DefEHreflected} respectively, and using the identities $\bm{M}^{\dagger}_{\bm{k}}\cdot\bm{M}_{\bm{k}} = \bm{N}^{\dagger}_{\bm{k}}\cdot\bm{N}_{\bm{k}} = 1$, $\bm{M}^{\dagger}_{\bm{k}}\cdot\bm{N}_{\bm{k}} = \bm{N}^{\dagger}_{\bm{k}}\cdot\bm{M}_{\bm{k}} = 0$, $\bm{M}^{\dagger}_{\bm{k}}\times\bm{M}_{\bm{k}} = \bm{N}^{\dagger}_{\bm{k}}\times\bm{N}_{\bm{k}} = 0$, $\bm{M}^{\dagger}_{\bm{k}}\times\bm{N}_{\bm{k}} = \bm{N}^{\dagger}_{\bm{k}}\times\bm{M}_{\bm{k}} = \ii\frac{c}{\omega}\bm{k}$ and $\bm{A}\cdot\left(\hat{z}\times\bm{B}\right) = \hat{z}\cdot\left(\bm{B}\times\bm{A}\right)$ one simplifies 
\begin{eqnarray}
\bm{M}^{\dagger}_{\bm{k}}\cdot\left(\hat{\bm{z}}\times[\bm{E}_{\text{in}} + \bm{E}_{\text{r}}]\right) & =  -\ii\frac{c}{\omega}k_{z}\left[A_{\parallel} - R_{\parallel}\right] & = - \hat{\bm{z}}\cdot\left(\bm{M}^{\dagger}_{\bm{k}}\times[\left(T_{o}\bm{E}_{\text{o}} + T_{e}\bm{E}_{\text{e}}\right)]\right)\\
\bm{N}^{\dagger}_{\bm{k}}\cdot\left(\hat{\bm{z}}\times[\bm{E}_{\text{in}} + \bm{E}_{\text{r}}]\right) & =  -\ii\frac{c}{\omega}k_{z}\left[A_{\perp} + R_{\perp}\right] & = - \hat{\bm{z}}\cdot\left(\bm{N}^{\dagger}_{\bm{k}}\times[\left(T_{o}\bm{E}_{\text{o}} + T_{e}\bm{E}_{\text{e}}\right)]\right)\nonumber\\
\bm{M}^{\dagger}_{\bm{k}}\cdot\left(\hat{\bm{z}}\times[\bm{H}_{\text{in}} + \bm{H}_{\text{r}} - ]\right) & =  \phantom{+\ii}\frac{c}{\omega}k_{z}\left[ R_{\perp} - A_{\perp} \right] & =  - \hat{\bm{z}}\cdot\left(\bm{M}^{\dagger}_{\bm{k}}\times[\left(T_{o}\bm{H}_{\text{o}} + T_{e}\bm{H}_{\text{e}}\right)]\right) + \frac{4\pi}{c}\bm{M}^{\dagger}_{\bm{k}}\cdot\sigma_{s}\left(T_{o}\bm{E}_{\text{o}} + T_{e}\bm{E}_{\text{e}}\right)\nonumber\\
\bm{N}^{\dagger}_{\bm{k}}\cdot\left(\hat{\bm{z}}\times[\bm{H}_{\text{in}} + \bm{H}_{\text{r}}]\right) & =  -\phantom{\ii}\frac{c}{\omega}k_{z}\left[A_{\parallel} + R_{\parallel}\right] & =  - \hat{\bm{z}}\cdot\left(\bm{N}^{\dagger}_{\bm{k}}\times[\left(T_{o}\bm{H}_{\text{o}} + T_{e}\bm{H}_{\text{e}}\right)]\right) + \frac{4\pi}{c}\bm{N}^{\dagger}_{\bm{k}}\cdot\sigma_{s}\left(T_{o}\bm{E}_{\text{o}} + T_{e}\bm{E}_{\text{e}}\right).\nonumber
\end{eqnarray}
The above system of equations is solved for 4 variables (reflection and transmission coefficients) as a function of the ($A_{\perp}$ and $A_{\parallel}$ variables, which are the amplitudes of the incident fields,
\begin{eqnarray}\label{FinalReflectionTransmissionMatrices}
R_{\perp} & = & \phantom{+}A_{\perp}\frac{(H^{M}_{e}-\sigma^{M}_{e}+ \ii  E^{N}_{e}) (E^{M}_{o}+ \ii  (H^{N}_{o}-\sigma^{N}_{o})) - (H^{M}_{o}-\sigma^{M}_{o}+ \ii  E^{N}_{o}) (E^{M}_{e}+ \ii    (H^{N}_{e}-\sigma^{N}_{e}))}{\Delta }\nonumber\\
& & + 2A_{\parallel}\frac{E^{N}_{e}(H^{M}_{o} -\sigma^{M}_{o}) - E^{N}_{o}(H^{M}_{e}-\sigma^{M}_{e})}{\Delta },\nonumber\\
R_{\parallel} & = & 2A_{\perp}\frac{E^{M}_{e} (H^{N}_{o} - \sigma^{N}_{o}) - E^{M}_{o} (H^{N}_{e} - \sigma^{N}_{e})}{\Delta}\nonumber\\
& & + A_{\parallel}\frac{(H^{M}_{e}-\sigma^{M}_{e}- \ii  E^{N}_{e}) (E^{M}_{o}- \ii  (H^{N}_{o}-\sigma^{N}_{o})) - (H^{M}_{o}-\sigma^{M}_{o}- \ii    E^{N}_{o}) (E^{M}_{e}- \ii  (H^{N}_{e}-\sigma^{N}_{e}))}{\Delta},\nonumber\\
T_{o} & = & \phantom{+}2A_{\perp}\frac{ c k_{z} (E^{M}_{e}+ \ii  (H^{N}_{e}-\sigma^{N}_{e}))}{\Delta\omega}
- 2 A_{\parallel}\frac{ c k_{z} (i   (H^{M}_{e}-\sigma^{M}_{e})+E^{N}_{e})}{\Delta\omega},\nonumber\\
T_{e} & = & - 2A_{\perp}\frac{ck_{z} (E^{M}_{o}+ \ii  (H^{N}_{o}-\sigma^{N}_{o}))}{\Delta\omega} + 2A_{\parallel}\frac{ ck_{z} (i (H^{M}_{o}-\sigma^{M}_{o})+E^{N}_{o})}{\Delta\omega}, \nonumber\\
\Delta & = & (H^{M}_{o}-\sigma^{M}_{o}- \ii  E^{N}_{o}) (E^{M}_{e}+ \ii  (H^{N}_{e}-\sigma^{N}_{e}))-(H^{M}_{e}-\sigma^{M}_{e}- \ii  E^{N}_{e}) (E^{M}_{o}+ \ii  (H^{N}_{o}-\sigma^{N}_{o})),
\end{eqnarray}
with the definitions $E^{M}_{o} = \bm{\tilde{M}}^{\dagger}_{\bm{k}}\cdot\bm{E}_{\text{o}}$, 
$E^{M}_{e} = \bm{\tilde{M}}^{\dagger}_{\bm{k}}\cdot\bm{E}_{\text{e}}$, $E^{N}_{o} = \bm{\tilde{N}}^{\dagger}_{\bm{k}}\cdot\bm{E}_{\text{o}}$, $E^{N}_{e} = \bm{\tilde{N}}^{\dagger}_{\bm{k}}\cdot\bm{E}_{\text{e}}$, $H^{M}_{o} = \bm{\tilde{M}}^{\dagger}_{\bm{k}}\cdot\bm{H}_{\text{o}}$, $H^{M}_{e} = \bm{\tilde{M}}^{\dagger}_{\bm{k}}\cdot\bm{H}_{\text{e}}$, $H^{N}_{o} = \bm{\tilde{N}}^{\dagger}_{\bm{k}}\cdot\bm{H}_{\text{o}}$, $H^{N}_{e} = \bm{\tilde{N}}^{\dagger}_{\bm{k}}\cdot\bm{H}_{\text{e}}$, $\sigma^{M}_{o} = \bm{M}^{\dagger}_{\bm{k},s}\cdot\bm{E}_{\text{o}}$, $\sigma^{M}_{e} = \bm{M}^{\dagger}_{\bm{k},s}\cdot\bm{E}_{\text{e}}$, $\sigma^{N}_{o} = \bm{N}^{\dagger}_{\bm{k},s}\cdot\bm{E}_{\text{o}}$ and $\sigma^{N}_{e} = \bm{N}^{\dagger}_{\bm{k},s}\cdot\bm{E}_{\text{e}}$. Also, the vectors $\bm{\tilde{M}}^{\dagger}_{\bm{k}}$, $\bm{\tilde{N}}^{\dagger}_{\bm{k}}$, $\bm{\tilde{M}}^{\dagger}_{\bm{k},s}$ and $\bm{\tilde{N}}^{\dagger}_{\bm{k},s}$ are defined as
\begin{eqnarray}
\bm{\tilde{M}}^{\dagger}_{\bm{k}} = \frac{-\ii}{\sqrt{ k_{x}^{2} + k_{y}^{2} }}\left(\begin{array}{c}
k_{x}\\
k_{y}\\
0
\end{array}
\right)e^{-\ii\left( \bm{k}\cdot\bm{r} - \omega t \right)},
\hspace{2cm}
\bm{M}^{\dagger}_{\bm{k},s} = \frac{4\pi}{c}\frac{\ii}{\sqrt{ k_{x}^{2} + k_{y}^{2} }}\left(\begin{array}{c}
k_{y}\sigma_{s,xx} - k_{x}\sigma_{s,yx}\\
k_{y}\sigma_{s,xy} - k_{x}\sigma_{s,yy}\\
0
\end{array}
\right)e^{-\ii\left( \bm{k}\cdot\bm{r} - \omega t \right)},
\end{eqnarray}
\begin{eqnarray}
\bm{\tilde{N}}^{\dagger}_{\bm{k}} = \frac{c}{\omega}\frac{k_{z}}{\sqrt{ k_{x}^{2} + k_{y}^{2} }}\left(\begin{array}{c}
k_{y}\\
- k_{x}\\
0
\end{array}
\right)e^{-\ii\left( \bm{k}\cdot\bm{r} - \omega t \right)},
\hspace{1.6cm}
\bm{N}^{\dagger}_{\bm{k},s} = \frac{4\pi}{c}\frac{c}{\omega}\frac{k_{z}}{\sqrt{ k_{x}^{2} + k_{y}^{2} }}\left(\begin{array}{c}
k_{x}\sigma_{xx} + k_{y}\sigma_{yx}\\
k_{x}\sigma_{xy} + k_{y}\sigma_{yy}\\
0
\end{array}
\right)e^{-\ii\left( \bm{k}\cdot\bm{r} - \omega t \right)}.
\end{eqnarray}
The above definitions are justified from the following relations,
\begin{eqnarray}
\hat{\bm{z}}\cdot\bm{M}^{\dagger}_{\bm{k}}\times\bm{A} & =
\left(\bm{z}\times\bm{M}^{\dagger}_{\bm{k}}\right)\cdot\bm{A} & =
\bm{\tilde{M}}^{\dagger}_{\bm{k}}\cdot\bm{A}\nonumber\\
\hat{\bm{z}}\cdot\bm{N}^{\dagger}_{\bm{k}}\times\bm{A} & =
\left(\bm{z}\times\bm{N}^{\dagger}_{\bm{k}}\right)\cdot\bm{A} & =
\bm{\tilde{N}}^{\dagger}_{\bm{k}}\cdot\bm{A},\nonumber\\
\frac{4\pi}{c}\left(\bm{M}^{\dagger}_{\bm{k}}\cdot\sigma_{s}\cdot\bm{A}\right) & =
\left(\frac{4\pi}{c}\bm{M}^{\dagger}_{\bm{k}}\cdot\sigma_{s}\right)\cdot\bm{A} & =
\bm{M}^{\dagger}_{\bm{k},s}\cdot\bm{A}\nonumber\\
\frac{4\pi}{c}\left(\bm{N}^{\dagger}_{\bm{k}}\cdot\sigma_{s}\cdot\bm{A}\right) & =
\left(\frac{4\pi}{c}\bm{N}^{\dagger}_{\bm{k}}\cdot\sigma_{s}\right)\cdot\bm{A} & =
\bm{N}^{\dagger}_{\bm{k},s}\cdot\bm{A}.
\end{eqnarray}
The reflection matrix elements $r_{MM},  r_{MN}, r_{NM}, r_{NN}$ can now be determined by comparing the matrix form in \Eq{DefinitionRT} and the relations for each component as given in \Eq{FinalReflectionTransmissionMatrices}. The full optical response of the WSM is now taken into account, such that the bulk components enter in the dielectric function (the components of the dielectric tensor are related to the components of the optical conductivity tensor via the relation $\epsilon_{ij}(\omega)=1-\frac{4\pi\ii \sigma_{ij}}{\omega}$, while the surface conductivity enters through the explicitly present $\sigma_{s,ij}$ in the boundary conditions of the magnetic field, in \Eq{EMBoundaryConditionPlate}. The so-obtained Fresnel coefficients are consistent with the Fresnel coefficients for anisotropic 3D Topological Insulators in \cite{Grushin2011} and for 2D Chern Insulators in \cite{RodriguezLopez2014}.

\newpage
\section{Reflection matrices for Weyl Semimetals}
We are interested in the Fresnel coefficients of the Weyl Semimetals. When $\bm{b} = (0,0,b)$, the electric permeability tensor is
\begin{eqnarray}
\bm{\epsilon} & = & \epsilon_{0}\left(\begin{array}{ccc}
\epsilon_{xx} & \epsilon_{xy} & 0\\
\epsilon_{yx} & \epsilon_{yy} & 0\\
0 & 0 & \epsilon_{zz}
\end{array}\right)
\end{eqnarray}
By using that $\bm{\epsilon}$ is hermitic, we have $\epsilon_{yx} = \epsilon_{xy}^{*}$. For purely imaginary frequency $\omega = \ii c\kappa$, all components of $\bm{\epsilon}$ are real, therefore, in this case, $\epsilon_{yx} = \epsilon_{xy}$.

The 2 different solutions obtained from the eigenproblem of $\bm{M}$ that goes to zero in the limit $z\to-\infty$ are
\begin{eqnarray}
E_{o,x} & = & \ii\left[k_{x}k_{y} \left(k_{x}^{2}\epsilon_{xx} + 2 k_{x} k_{y}\epsilon_{xy} + k_{y}^{2}\epsilon_{yy} - \epsilon_{zz}q_{e}^{2}\right) + \mu\epsilon_{zz}\kappa^{2}\left( \epsilon_{xy}k_{\parallel}^{2} +k_{x} k_{y} (\epsilon_{xx}+\epsilon_{yy}-\epsilon_{zz}) \right) + \epsilon_{xy}\mu^{2}\epsilon_{zz}^{2}\kappa^{4}\right]\nonumber\\
E_{o,y} & = & \ii\left[ k_{x}^{2}\left( \epsilon_{xx}k_{y}^{2} + \mu\epsilon_{zz}^{2}\kappa^{2} \right)+\left( k_{y}^{2} + \mu\epsilon_{zz}\kappa^{2} \right)\left( 2 k_{x} k_{y} \epsilon_{xy} + k_{y}^{2} \epsilon_{yy} + \epsilon_{zz}\left( \mu\epsilon_{yy}\kappa^{2} - q_{e}^{2} \right) \right)\right]\nonumber\\
H_{o,x} & = & - \ii\kappa  q_{o} \epsilon_{zz} \left(k_{x}^2 \epsilon_{zz}+k_{x} k_{y} \epsilon_{xy}+k_{y}^2 \epsilon_{yy} + \epsilon_{zz}\left( \mu\epsilon_{yy}\kappa^{2} - q_{e}^{2} \right)\right)\nonumber\\
H_{o,y} & = & \ii\kappa  q_{o} \epsilon_{zz} \left(k_{x} k_{y} (\epsilon_{xx}-\epsilon_{zz})+\epsilon_{xy} \left( k_{y}^{2} + \mu\epsilon_{zz}\kappa^{2} \right)\right)
\end{eqnarray}

\begin{eqnarray}
E_{e,x} & = & \ii\left[ k_{x} k_{y}\left( k_{x}^{2}\epsilon_{xx} + 2 k_{x} k_{y}\epsilon_{xy} + k_{y}^{2}\epsilon_{yy} - \epsilon_{zz}q_{o}^{2} \right) + \mu\epsilon_{zz}\kappa^{2}\left( \epsilon_{xy}k_{\parallel}^{2} +k_{x} k_{y} (\epsilon_{xx}+\epsilon_{yy}-\epsilon_{zz}) \right) + \epsilon_{xy}\mu^{2}\epsilon_{zz}^{2}\kappa^{4}\right]\nonumber\\
E_{e,y} & = & \ii\left[ k_{x}^{2}\left( \epsilon_{xx}k_{y}^{2} + \mu\epsilon_{zz}^{2}\kappa^{2} \right)+\left( k_{y}^{2} + \mu\epsilon_{zz}\kappa^{2} \right)\left( 2 k_{x} k_{y} \epsilon_{xy} + k_{y}^{2} \epsilon_{yy} + \epsilon_{zz}\left( \mu\epsilon_{yy}\kappa^{2} - q_{o}^{2} \right) \right)\right]\nonumber\\
H_{e,x} & = & - \ii\kappa  q_{e} \epsilon_{zz} \left(k_{x}^2 \epsilon_{zz}+k_{x} k_{y} \epsilon_{xy}+k_{y}^2 \epsilon_{yy} + \epsilon_{zz}\left( \mu\epsilon_{yy}\kappa^{2} - q_{o}^{2} \right)\right)\nonumber\\
H_{e,y} & = & \ii\kappa  q_{e} \epsilon_{zz} \left(k_{x} k_{y} (\epsilon_{xx}-\epsilon_{zz})+\epsilon_{xy} \left( k_{y}^{2} + \mu\epsilon_{zz}\kappa^{2} \right)\right)
\end{eqnarray}
with eigenvalues $q_{o} = -\ii\sqrt{ S + \sqrt{S^{2} - T} }$ and $q_{e} = -\ii\sqrt{ S - \sqrt{S^{2} - T} }$, where
\begin{eqnarray}
S & = & \frac{k_{x}^2 (\epsilon_{xx}+\epsilon_{zz})+2 k_{x} k_{y} \epsilon_{xy}+k_{y}^2 (\epsilon_{yy}+\epsilon_{zz})+\kappa^{2}\mu  \epsilon_{zz} (\epsilon_{xx}+\epsilon_{yy})}{2 \epsilon_{zz}}\\
T & = & \frac{\left(k_{\parallel}^{2} + \mu\epsilon_{zz}\kappa^{2} \right) \left(k_{x}^2 \epsilon_{xx}+2 k_{x} k_{y} \epsilon_{xy}+k_{y}^2 \epsilon_{yy}+\kappa^{2}\mu  \left(\epsilon_{xx} \epsilon_{yy}-\epsilon_{xy}^2\right)\right)}{\epsilon_{zz}}
\end{eqnarray}
\subsection{Reflection matrix with surface conductivities}
Using the formulas derived above, the Fresnel reflection matrix is
\begin{eqnarray}
\mathbb{R} & = & \left(\begin{array}{cc}
r_{MM} & r_{MN}\\
r_{NM} & r_{NN}
\end{array}\right) = \frac{1}{\Delta}\left(\begin{array}{cc}
R_{MM} & R_{MN}\\
R_{NM} & R_{NN}
\end{array}\right),
\end{eqnarray}
where
\begin{eqnarray}
R_{MM} & = & \epsilon_{zz}R_{MM,0} - \frac{2}{\kappa}R_{MM,1} - D\sigma,\nonumber\\
R_{MN} & = & \epsilon_{zz}\left( R_{MN,0} + 2R_{MN,1} \right),\nonumber\\
R_{NM} & = & \epsilon_{zz}\left( R_{NM,0} + 2\mu R_{NM,1} \right),\nonumber\\
R_{NN} & = & \epsilon_{zz}R_{NN,0} + \frac{2}{\kappa}R_{NN,1} + D\sigma,\nonumber\\
\Delta  & = &  \epsilon_{zz}\Delta_{0} + \frac{2}{\kappa}\Delta_{1} + D\sigma,
\end{eqnarray}
where the first term of each component is independent of the surface conductivities, the second term is linear in surface conductivities terms, and the third term is proportional to the determinant of the surface conductivity tensor, with
\begin{eqnarray}
R_{MM,0} & = & I_{e}F_{o}( q_{e} - \mu\kappa_{z} )\left( k_{\parallel}^{2}+ \epsilon_{zz}\left( \mu\kappa^{2} + q_{o}\kappa_{z} \right)\right) - I_{o}F_{e} (q_{o} - \mu\kappa_{z} ) \left( k_{\parallel}^{2} + \epsilon_{zz}\left( \mu\kappa^{2} + q_{e}\kappa_{z} \right)\right)\\
R_{MN,0} & = & 2\ii\kappa_{z}\mu(q_{e}-q_{o}) \kappa F_{o}F_{e}\\
R_{NM,0} & = & -2\ii\kappa_{z}\frac{\epsilon_{zz}}{\kappa}\left(k_{\parallel}^{2} + \mu\epsilon_{zz}\kappa ^{2} \right)(q_{e}-q_{o}) I_{o}I_{e}\\
R_{NN,0} & = & I_{e}F_{o}( q_{e} + \mu\kappa_{z} ) \left( k_{\parallel}^{2} + \epsilon_{zz}\left( \mu  \kappa^{2} - q_{o}\kappa_{z} \right)\right) - I_{o}F_{e} ( q_{o} + \mu\kappa_{z} ) \left( k_{\parallel}^{2} + \epsilon_{zz}\left( \mu\kappa^{2} - q_{e}\kappa_{z} \right)\right)\\
\Delta_{0} & = & I_{o}F_{e}( q_{o} + \mu\kappa_{z} )\left( k_{\parallel}^{2} + \epsilon_{zz}\left( \mu  \kappa^{2} + q_{e}\kappa_{z} \right)\right) - I_{e}F_{o} ( q_{e} + \mu\kappa_{z} ) \left( k_{\parallel}^{2} +\epsilon_{zz}\left( \mu\kappa^{2} + q_{o}\kappa_{z} \right)\right),\\
R_{MM,1} & = &
\kappa_{z}F_{o}( \mu\kappa_{z} - q_{e} ) (G_{e} (\tilde{\sigma}_{xx} k_{x}+\tilde{\sigma}_{yx} k_{y})+C_{e} (\tilde{\sigma}_{xy} k_{x}+\tilde{\sigma}_{yy} k_{y}))
\nonumber\\& &
 - \kappa_{z}F_{e}( \mu\kappa_{z} - q_{o} ) (G_{o} (\tilde{\sigma}_{xx} k_{x}+\tilde{\sigma}_{yx} k_{y})+C_{o} (\tilde{\sigma}_{xy} k_{x}+\tilde{\sigma}_{yy} k_{y}))
\nonumber\\& &
 - I_{o}\left(k_{\parallel}^2+\epsilon_{zz} \left(\kappa ^{2} \mu +\kappa_{z} q_{e}\right)\right) (G_{e} (\tilde{\sigma}_{xx} k_{y}-\tilde{\sigma}_{yx} k_{x})+C_{e} (\tilde{\sigma}_{xy} k_{y}-\tilde{\sigma}_{yy} k_{x}))
\nonumber\\& &
 + I_{e}\left(k_{\parallel}^2+\epsilon_{zz} \left(\kappa ^{2} \mu +\kappa_{z} q_{o}\right)\right) (G_{o} (\tilde{\sigma}_{xx} k_{y}-\tilde{\sigma}_{yx} k_{x})+C_{o} (\tilde{\sigma}_{xy} k_{y}-\tilde{\sigma}_{yy} k_{x}))\\
R_{MN,1} & = & 2\ii\kappa_{z}\left( q_{e}^{2} - q_{o}^{2} \right) \left( k_{\parallel}^{2} + \mu\epsilon_{zz}\kappa^{2} \right) (k_{x} (\tilde{\sigma}_{xx} k_{y} - \tilde{\sigma}_{yx} k_{x}) + k_{y} (\tilde{\sigma}_{xy} k_{y} - \tilde{\sigma}_{yy} k_{x}))\nonumber\\
& & \times \left( k_{x} k_{y} (\epsilon_{xx}-\epsilon_{zz})+\epsilon_{xy} \left(k_{y}^{2} + \mu\epsilon_{zz}\kappa^{2} \right)\right)\\
R_{NM,1} & = & 2\ii\kappa_{z}\mu\left(q_{e}^2-q_{o}^2\right)\left( k_{\parallel}^{2} + \mu\epsilon_{zz}\kappa^{2} \right) (k_{y} (\tilde{\sigma}_{xx} k_{x}+\tilde{\sigma}_{yx} k_{y})-k_{x} (\tilde{\sigma}_{xy} k_{x}+\tilde{\sigma}_{yy} k_{y}))\nonumber\\
& & \times \left(k_{x} k_{y} (\epsilon_{xx}-\epsilon_{zz})+\epsilon_{xy} \left(k_{y}^{2} + \mu\epsilon_{zz}\kappa^{2} \right)\right)\\
R_{NN,1} & = &
- \kappa_{z}F_{o}( \mu\kappa_{z} + q_{e} ) (G_{e} (\tilde{\sigma}_{xx} k_{x}+\tilde{\sigma}_{yx} k_{y})+C_{e} (\tilde{\sigma}_{xy} k_{x}+\tilde{\sigma}_{yy} k_{y}))
\nonumber\\& &
+ \kappa_{z}F_{e}( \mu\kappa_{z} + q_{o} ) (G_{o} (\tilde{\sigma}_{xx} k_{x}+\tilde{\sigma}_{yx} k_{y})+C_{o} (\tilde{\sigma}_{xy} k_{x}+\tilde{\sigma}_{yy} k_{y}))
\nonumber\\& &
- I_{o}\left(\kappa_{z} q_{e} \epsilon_{zz}-\left(k_{\parallel}^2+\kappa ^{2} \mu  \epsilon_{zz}\right)\right) (G_{e} (\tilde{\sigma}_{xx} k_{y}-\tilde{\sigma}_{yx} k_{x})+C_{e} (\tilde{\sigma}_{xy} k_{y}-\tilde{\sigma}_{yy} k_{x}))
\nonumber\\& &
+ I_{e}\left(\kappa_{z} q_{o} \epsilon_{zz}-\left(k_{\parallel}^2+\kappa ^{2} \mu  \epsilon_{zz}\right)\right) (G_{o} (\tilde{\sigma}_{xx} k_{y}-\tilde{\sigma}_{yx} k_{x})+C_{o} (\tilde{\sigma}_{xy} k_{y}-\tilde{\sigma}_{yy} k_{x}))\\
\Delta_{1} & = & 
- \kappa_{z}F_{o}( \mu\kappa_{z} + q_{e} ) (G_{e} (\tilde{\sigma}_{xx} k_{x}+\tilde{\sigma}_{yx} k_{y})+C_{e} (\tilde{\sigma}_{xy} k_{x}+\tilde{\sigma}_{yy} k_{y}))
\nonumber\\& &
+ \kappa_{z}F_{e}( \mu\kappa_{z} + q_{o} ) (G_{o} (\tilde{\sigma}_{xx} k_{x}+\tilde{\sigma}_{yx} k_{y})+C_{o} (\tilde{\sigma}_{xy} k_{x}+\tilde{\sigma}_{yy} k_{y}))
\nonumber\\& &
- I_{o}\left(k_{\parallel}^2+\epsilon_{zz} \left(\kappa ^{2}\mu +\kappa_{z} q_{e}\right)\right) (G_{e} (\tilde{\sigma}_{xx} k_{y}-\tilde{\sigma}_{yx} k_{x})+C_{e} (\tilde{\sigma}_{xy} k_{y}-\tilde{\sigma}_{yy} k_{x}))
\nonumber\\& &
+ I_{e}\left(k_{\parallel}^2+\epsilon_{zz} \left(\kappa ^{2}\mu +\kappa_{z} q_{o}\right)\right) (G_{o} (\tilde{\sigma}_{xx} k_{y}-\tilde{\sigma}_{yx} k_{x})+C_{o} (\tilde{\sigma}_{xy} k_{y}-\tilde{\sigma}_{yy} k_{x}))\\
D\sigma & = & 4 \kappa_{z} k_{\parallel}^{2}\mu\epsilon_{zz}\left(q_{o}^2-q_{e}^2\right)  \left(k_{\parallel}^2+\mu\epsilon_{zz}\kappa^{2}\right) \left(k_{x} k_{y} (\epsilon_{xx}-\epsilon_{zz})+\epsilon_{xy} \left(k_{y}^2+\mu\epsilon_{zz}\kappa^{2}\right)\right)(\tilde{\sigma}_{xx} \tilde{\sigma}_{yy}-\tilde{\sigma}_{xy} \tilde{\sigma}_{yx}),
\end{eqnarray}
where we have used $\tilde{\sigma}_{ij} = \frac{2\pi}{c}\sigma_{ij}$, $k_{z} = \ii\kappa_{z}$ (therefore, $\kappa_{z} = \sqrt{ k_{\parallel}^{2} + \kappa^{2} }$), $k_{\parallel} = \sqrt{ k_{x}^{2} + k_{y}^{2} }$, and
\begin{eqnarray}
C_{o} & = & \left( k_{y}^{2} + \mu\epsilon_{zz}\kappa^{2} \right)\left( 2\epsilon_{xy}k_{x}k_{y} + \epsilon_{yy}k_{y}^{2} + \epsilon_{zz}\left( \mu\epsilon_{yy}\kappa^{2} - q_{o}^{2} \right)\right) + k_{x}^{2}\left( \epsilon_{xx}k_{y}^{2} + \mu\epsilon_{zz}^{2}\kappa^{2} \right)\nonumber\\
C_{e} & = & \left( k_{y}^2 + \mu\epsilon_{zz}\kappa^{2} \right)\left( 2\epsilon_{xy}k_{x}k_{y} + \epsilon_{yy}k_{y}^{2} + \epsilon_{zz}\left( \mu\epsilon_{yy}\kappa^{2} - q_{e}^{2} \right)\right) + k_{x}^{2}\left( \epsilon_{xx}k_{y}^{2} + \mu\epsilon_{zz}^{2}\kappa^{2} \right)\nonumber\\
F_{o} & = & k_{x} \left(\epsilon_{zz} \left(k_{\parallel}^2-q_{o}^2+\mu\epsilon_{yy}\kappa^{2}\right)+k_{y}^{2}(\epsilon_{yy}-\epsilon_{xx})\right)-k_{y} \epsilon_{xy} \left(k_{y}^{2} - k_{x}^{2}+\mu\epsilon_{zz}\kappa^{2}\right)\nonumber\\
F_{e} & = & k_{x} \left(\epsilon_{zz} \left(k_{\parallel}^2-q_{e}^2+\mu\epsilon_{yy}\kappa^{2}\right)+k_{y}^{2}(\epsilon_{yy}-\epsilon_{xx})\right)-k_{y} \epsilon_{xy} \left(k_{y}^{2} - k_{x}^{2}+\mu\epsilon_{zz}\kappa^{2}\right)\nonumber\\
G_{o} & = & \mu\epsilon_{zz}\kappa^{2} \left(\epsilon_{xy} \left(k_{\parallel}^2+\mu\epsilon_{zz}\kappa^{2}\right)+k_{x} k_{y} (\epsilon_{xx}+\epsilon_{yy}-\epsilon_{zz})\right)+k_{x} k_{y} \left(k_{x}^{2}\epsilon_{xx}+2 k_{x} k_{y} \epsilon_{xy}+k_{y}^{2}\epsilon_{yy}-q_{o}^{2}\epsilon_{zz}\right)\nonumber\\
G_{e} & = & \mu\epsilon_{zz}\kappa^{2} \left(\epsilon_{xy} \left(k_{\parallel}^2+\mu\epsilon_{zz}\kappa^{2}\right)+k_{x} k_{y} (\epsilon_{xx}+\epsilon_{yy}-\epsilon_{zz})\right)+k_{x} k_{y} \left(k_{x}^{2}\epsilon_{xx}+2 k_{x} k_{y} \epsilon_{xy}+k_{y}^{2}\epsilon_{yy}-q_{e}^{2}\epsilon_{zz}\right)\nonumber\\
I_{o} & = & \frac{k_{y}}{\epsilon_{z}}\left(k_{x}^{2}\epsilon_{xx}+2 k_{x} k_{y} \epsilon_{xy}+k_{y}^{2}\epsilon_{yy}-q_{o}^{2}\epsilon_{zz}\right)+\mu\kappa^{2} (k_{x} \epsilon_{xy}+k_{y} \epsilon_{yy})\nonumber\\
I_{e} & = & \frac{k_{y}}{\epsilon_{z}}\left(k_{x}^{2}\epsilon_{xx}+2 k_{x} k_{y} \epsilon_{xy}+k_{y}^{2}\epsilon_{yy}-q_{e}^{2}\epsilon_{zz}\right)+\mu\kappa^{2} (k_{x} \epsilon_{xy}+k_{y} \epsilon_{yy})
\end{eqnarray}

It is easy to check that $r_{NM} = r_{MN}$.
\subsection{Reflection matrix without surface conductivities}
Using the formulas derived above, the Fresnel reflection matrix (in the absence of surface conductivities) is
\begin{eqnarray}
\mathbb{R} & = & \left(\begin{array}{cc}
r_{MM} & r_{MN}\\
r_{NM} & r_{NN}
\end{array}\right) = \frac{1}{\Delta_{0}}\left(\begin{array}{cc}
R_{MM,0} & R_{MN,0}\\
R_{NM,0} & R_{NN,0}
\end{array}\right).
\end{eqnarray}

It is easy to check that $r_{NM} = r_{MN}$.

\subsection{Limit of small Hall conductivity}
Applying a Taylor expansion to the result from Eqs. (64) with respect to the Hall conductivity being small as compared to the diagonal components, one obtaines
\begin{eqnarray}
\lim_{\epsilon_{xy}\ll 1}\mathbb{R} & = & \left(\begin{array}{cc}
\frac{\mu\kappa_{z} - q_{1}}{\mu\kappa_{z} + q_{1}} & 0\\
0 & \frac{\epsilon\kappa_{z} - q_{1}}{\epsilon\kappa_{z} + q_{2}}
\end{array}
\right) + 2\epsilon_{xy}\frac{\kappa_{z}}{k_{\parallel}^{2}}\left(\begin{array}{cc}
\frac{k_{x}k_{y}\kappa^{2}\mu^{2}}{q_{1}\left(\mu\kappa_{z} + q_{1}\right)^{2}} & \ii\frac{\left(k_{x}^{2} - k_{y}^{2}\right)\mu\epsilon_{zz}\kappa q_{2}\left( q_{1} - q_{2} \right)}{k_{\parallel}^{2}\left( \epsilon - \epsilon_{zz} \right)\left( \mu\kappa_{z} + q_{1} \right)\left( \epsilon\kappa_{z} + q_{2} \right)}\\
\ii\frac{\left(k_{x}^{2} - k_{y}^{2}\right)\mu\epsilon_{zz}\kappa q_{2}\left( q_{1} - q_{2} \right)}{k_{\parallel}^{2}\left( \epsilon - \epsilon_{zz} \right)\left( \mu\kappa_{z} + q_{1} \right)\left( \epsilon\kappa_{z} + q_{2} \right)} & \frac{k_{x}k_{y}q_{2}}{\left(\epsilon\kappa_{z} + q_{2}\right)^{2}}
\end{array}
\right) + \mathcal{O}\left[ \epsilon_{xy}^{2} \right],\nonumber\\
\end{eqnarray}
with $q_{1} = \sqrt{ k_{\parallel}^{2} + \mu\epsilon\kappa^{2} }$ and $q_{2} = \sqrt{ \frac{\epsilon}{\epsilon_{zz}}k_{\parallel}^{2} + \mu\epsilon\kappa^{2} }$. When $\mu = \epsilon_{xx} = \epsilon_{yy} = \epsilon_{zz}= 1$, for small $\epsilon_{xy}$, the R matrix is approximated by
\begin{eqnarray}
\lim_{\epsilon_{xy} \ll 1}\mathbb{R} & = & \frac{\epsilon_{xy}}{2k_{\parallel}^{2}}\left(\begin{array}{cc}
k_{x}k_{y}\left(\frac{\kappa}{k_{z}}\right)^{2} & \ii\frac{k_{y}^{2} - k_{x}^{2}}{2}\frac{\kappa}{k_{z}} \\
- \ii\frac{k_{y}^{2} - k_{x}^{2}}{2}\frac{\kappa}{k_{z}} & k_{x}k_{y}\\
\end{array}\right) + \mathcal{O}\left[\epsilon_{xy}^{2}\right],
\end{eqnarray}
